\documentclass[11pt,a4paper,sn-basic]{article}
\usepackage[utf8]{inputenc}
\usepackage{amsmath, amssymb}
\usepackage{float}
\usepackage{graphicx, todonotes}
\usepackage{authblk}
\usepackage{amsthm,url,fullpage}
\usepackage[toc,page]{appendix}
\theoremstyle{definition} 

\usepackage{comment}
\usepackage{multirow,subfig}
\theoremstyle{remark}
\usepackage{booktabs}

\title{Quadratic and Higher-Order Unconstrained Binary Optimization of Railway Rescheduling for Quantum Computing}

\author[1]{Krzysztof Domino \thanks{kdomino@iitis.pl, ORCID: 0000-0001-7386-5441}}
\author[1,3]{Akash Kundu \thanks{akundu@iitis.pl, ORCID: 0000-0002-3540-1061}}
\author[1]{\"Ozlem Salehi \thanks{osalehi@iitis.pl, ORCID: 0000-0003-2033-2881}}
\author[2]{Krzysztof Krawiec \thanks{krzysztof.krawiec@polsl.pl, ORCID: 0000-0002-7447-3447}}

\affil[1]{Institute of Theoretical and Applied Informatics\\ 
	Polish Academy of Sciences\\ 
	Ba{\l}tycka~5, 44-100 Gliwice, Poland}
\affil[2]{Silesian University of Technology, Faculty of Transport and Aviation Engineering, Akademicka 2A, 44-100 Gliwice, Poland} 
\affil[3]{Joint Doctoral School, Silesian University of Technology, Akademicka 2A, 44-100 Gliwice, Poland}
\begin{document}

\maketitle

\section*{Abstract}

As consequences of disruptions in railway traffic affect passenger experience/satisfaction, appropriate rerouting and/or rescheduling is necessary. These problems are known to be NP-hard, given the numerous restrictions of traffic nature. With the recent advances in quantum technologies, quantum annealing has become an alternative method to solve such optimization problems. To use quantum annealing, the problem needs to be encoded in QUBO (quadratic unconstrained binary optimization) or HOBO  (higher-order binary optimization) formulation that can be recast as a QUBO. This paper introduces QUBO and HOBO representations for rescheduling problems of railway traffic management; the latter is a new approach up to our knowledge. This new approach takes into account not only the single-track lines but also the double- and multi-track lines, as well as stations composed of tracks and switches. We consider the conditions of minimal headway between trains, minimal stay on stations, track occupation, and rolling stock circulation. Furthermore, a hybrid quantum-classical procedure is presented that includes rerouting. We demonstrate the proof of concept implementation on the D-Wave Quantum Processing Unit and D-Wave hybrid solver. 

\section*{Keywords}
higher-order binary optimization, railway rescheduling, railway rerouting, quantum-classical hybrid procedure, D-Wave annealer.

\section{Introduction}

Railway transport is perceived as a more sustainable and ecological alternative to individual mobility~\cite{Mulley2021, BATTY2015109}. The increasing train traffic and other safety-related issues cause dispatching problems in case of disturbances which may lead to rerouting and rescheduling. Failure to resolve them quickly and efficiently can cause inconvenience for the passengers and increase the costs. No matter what the reason for the disturbance is (technical malfunction of traffic control, collision with car or animal, system -- see \cite{gawlak_2022}), the objective is to reduce delay propagation~\cite{Larsen2014, TORNQUIST2007342, Corman2011}.

One can harness quantum computing for solving the railway rescheduling and rerouting problem formulated as an optimization problem. A promising heuristic algorithm is quantum annealing, which relies on the quantum adiabatic model of computation \cite{apolloni1989quantum, kadowaki1998quantum, farhi2000quantum}. Commercially available quantum annealers are provided by the D-Wave company \cite{johnson2011quantum}. The problem of interest needs to be formulated as an Ising model, which in turn determines the coupling strength between the pair of qubits of the machine. After the system is evolved slowly enough for a particular duration, it is expected to be found in the minimum energy state, encoding the solution that minimizes the objective function of the problem. Any problem that is formulated as a quadratic unconstrained binary optimization (QUBO) can be easily transformed into an Ising model and solved using quantum annealing in principle. Since it is more natural to express problems using QUBO representation than the Ising model, it is desirable to find QUBO formulations for optimization problems to target quantum annealing \cite{glover2018tutorial, lucas2014ising}. A generalization of QUBO is higher-order binary optimization (HOBO) representation that allows not only quadratic terms but also higher-order terms in the objective function. There has been some recent work on formulating HOBO representations for combinatorial optimization problems in the context of quantum optimization \cite{glos2020space, tabi2020quantum, salehi2021unconstrained}.

The motivation of this paper is to demonstrate that it is possible to encode typical railway infrastructure and traffic conditions as QUBOs and HOBOs, making the problems quantum computing ready. This paper is a follow-up of~\cite{domino2020quantum} which comprises railway rescheduling under disruptions on a single-track railway line encoded using QUBO. Now we remove the restriction of single-track lines, enabling also double- and multi-track lines on model trains traffic on stations. We use a parallel machine approach improved by rerouting, resulting in a hybrid algorithm. The presented representations for railway rescheduling and rerouting include the conditions of the minimal headway between trains, minimal stay on stations, station/track occupation, and rolling stock circulation. We use a classical procedure that mimics real-life rerouting practices together with quantum annealing to solve the rescheduling problem and end up with a hybrid algorithm. Although the detailed discussion of our approach concerns the railway rescheduling problem introduced here, similar approaches can be adopted for problems from other branches of operational research such as factory trolleys or electric busses rescheduling/rerouting.

There is a vast amount of research in the scope of resuming the railway system's capacity and proper functioning after a disruption; for a systematic review see ~\cite{MATTSSON201516}. There are also publications in which other techniques like genetic algorithms and deep learning techniques are used. One may find out more in numerous review papers \cite{8795577, Cordeau1998, 6920082} in the scope of optimization methods to solve railway conflict management problems. Given the NP-hardness of such rescheduling problems and their complexity, it is very challenging to solve them on current computational devices in a reasonable time. We expect quantum computing to offer novel opportunities to overcome these limitations.

In our approach, we chose the parallel machine approach, where trains have a fixed route within the stations~\cite{LIU20092840}. The reason is that passenger trains have fixed platforms within the station, and the platform change is an extraordinary situation that affects passengers. For demonstration reasons, we start with an Integer Linear Programming formulation where we use order variables~\cite{5874901} to determine the order of trains leaving the station. Alternatively, for the QUBO and HOBO approaches, we use discrete-time units~\cite{HARROD2009830}, in which binary variables describe whether the event happens at a given time. 

Our paper follows other research efforts towards solving transportation-related problems using quantum annealers~\cite{salehi2021unconstrained, borowski2020new, neukart2017traffic} or quantum approximate optimization algorithm~\cite{farhi2014quantum} (QAOA)~\cite{glos2020space}. HOBOs are considered in some mentioned papers for various transportation problems. However, up to our knowledge, HOBO formulation is considered to address the railway rescheduling problem for the first time.

The paper is organized as follows. Section \ref{s::RailwaySystemModel} gives a brief overview of the railway system model, which consists of infrastructure and traffic. In this section, we present the notions and formalism to describe the problem of railways rescheduling. In Section \ref{sec:prob} we present a linear programming representation, we set out the QUBO and HOBO formulations, and we describe our approach to rerouting. We demonstrate the formulations in Section \ref{s::Demonstration} both theoretically and using numerical calculations.
The last section contains conclusions and a discussion on the possibility of further development of QUBO and HOBO representations to address railway rescheduling. 

\section{Railway system model}\label{s::RailwaySystemModel}

Trains run according to a \emph{schedule} along the routes. The route of the train is composed of stations and lines between them.
The line consists of one or more parallel tracks, each split into \emph{line blocks}. The latter we understand as a track section between two signaling utilities that can only be occupied by one train at a time. Stations consist of tracks interconnected by railroad switches (referred to as \emph{switches}). Similar to \emph{line blocks}, stations consist of \emph{station blocks} -- track sections at stations between two signaling utilities that can be occupied by only one train at a time. Trains are controlled by dispatchers who can reroute or/and reschedule them if necessary. By rerouting we understand the change of the track used by a train within a line or a station. By rescheduling, we understand the modification of the train departure time in a way to avoid conflict and maintain the feasibility of the timetable. (Note that we define infrastructure terms from train traffic perspective rather than their physical characteristics, which is not the usual description in transportation research -- we keep this description to keep it coherent with our mathematical model aiming to make it more illustrative.)

Two trains meet and pass (M-P) meeting at the same spatial location while following the same route in opposite directions. Similarly, two trains meet and overtake (M-O) when one train overtakes another. Depending on the type of railway line, M-Ps and M-Os may occur at stations and/or on lines. We distinguish single-track, double-track, and multi-track lines. On \emph{single-track} lines, trains can M-P and M-O only at stations. The usual use of \emph{double-track} line is such that trains are heading in one direction on one track and in the other direction on the other track (unidirectional traffic). It implies M-P possibility at stations and lines and M-O possibility only at stations. We also consider another use of double-track lines as two parallel single-track lines (bidirectional traffic). In this mode, trains can M-O on the line between stations while heading in the same direction on both tracks (this is at the cost of M-P possibility). The bidirectional mode may also be used on \emph{multi-track lines}.

Regardless of the type of line, trains need to keep minimal headway -- the distance between two trains following the same direction to preserve safety conditions. Such headway can be measured either in space or in time if taking into account trains' speeds.
Trains can terminate at a station, set off from a station, have scheduled stop there, or pass it by.
As \emph{conflict}~\cite{d2007branch} we understand the situation that occurs when at least two trains compete for the same resource in terms of blocks (station or line) or switches. In our model, we aim to resolve optimally all conflicts by rescheduling and rerouting while keeping the safety conditions and limiting the schedule modification.

For model's simplicity, let us assume that the schedule is a pre-set sequences of blocks with departure times assigned. We will refer to this as the \emph{default settings}; any change will be considered as rerouting and creation the new model. In real rail traffic, the schedule is assumed to be conflict free, and conflicts appear due to \emph{delays}. We define \emph{delays} as the difference between $t(j, s_{in})$ or $t(j, s_{out})$ -- the actual time of entering or leaving particular station $s$ by train $j$, and the scheduled time $\sigma(j,s_{in})$ or $\sigma(j,s_{out})$. In the rest of this section we use $s^*$ for either $s_{in}$ or $s_{out}$. Hence, the delay is:
\begin{equation}\label{eq::delay_repr}
d(j,s^*) = t(j,s^*) - \sigma(j,s^*).
\end{equation}

Following~\cite{d2007branch}, we split the delay into \emph{unavoidable} $d_u(j,s^*)$ and \emph{additional} $d_a(j,s^*)$ in the following manner: \begin{equation}\label{eq::split_d}
d(j,s^*) = d_u(j,s^*) + d_a(j,s^*).
\end{equation} By unavoidable delay, we understand the delay from outside the model that is propagated through the network, not including any delay that may be caused by other trains' traffic and that can not be controlled in the model. (Unavoidable delay may be caused by accidents, technical failure, delay from outside the analyzed network, or delays of the trains affected by those on subsequent stations.) The additional delay comprises delays beyond unavoidable caused by solving conflicts due to traffic, which is in control of our model. The latter is of main interest to us as our goal is to minimize the additional delays.

As we intend not to extend the delays, we assume that the additional delays are limited by the parameter $d_{\text{max}}$:
\begin{equation}\label{eq::d_limit}
0 \leq d_a(j,s^*) \leq d_{\text{max}}.
\end{equation}
$d_{\text{max}}$ is a parameter of the model and limits the range of the integer variables in the linear model and the number of variables in QUBO or HOBO approaches; as such, it affects the problem size. It should not be set too low, resulting in a situation in which obtaining a feasible solution is not possible. There are a few possibilities for determining $d_{\text{max}}$. Following ~\cite{domino2020quantum} and Tab. $1$ therein, one can use some simple heuristics such as FCFS (first come first serve) or FLFS (first leave first serve) to get the solution that is not optimal but feasible. (As discussed in~\cite{d2007branch} these heuristics are often used in real live railway rescheduling.) Such simple heuristics and solutions can be used to determine $d_{\max}$ for the practical problem. (Bear in mind that in an advanced model, $d_{\max}$ may also be train and station dependent.)

\begin{figure}[tbh!]
\centering
\includegraphics[width = \linewidth]{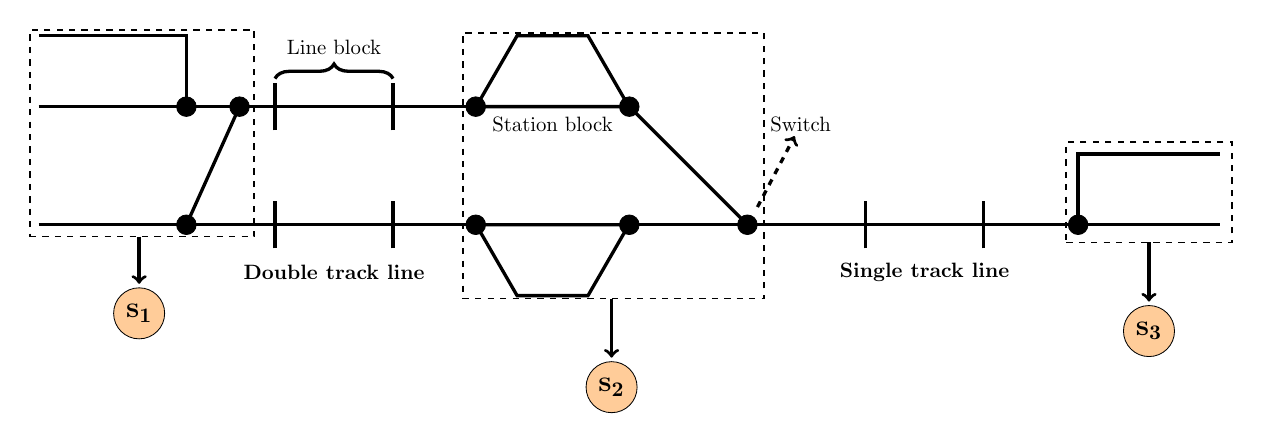}
\caption{A comprehensive yet simplified illustration of railway infrastructure.}
\label{fig::railway-infrastructure}
\end{figure}

A summary of definition of railway terminologies is given in Tab.~\ref{tab:railways-terminology-definition} in Appendix \ref{appendix:: definitions-rail-term}. The comprehensive illustration of railway infrastructure is given in Fig.~\ref{fig::railway-infrastructure}.

\section{Problem formulation}\label{sec:prob}

In this section, we discuss the conditions that need to be satisfied and the objective function of the problem. The symbols used are summarised in Tab.~\ref{tab::symbols}.
Following~\cite{domino2020quantum}, our goal is to minimize the weighted additional delay
\begin{equation}\label{eq::fd_without_dmax}
    f_o = \sum_{j \in \mathcal{J}} w_j \sum_{s \in \mathcal{S}_j} d_a(j, s_{\text{out}}),
\end{equation}
where $\mathcal{J}$ is the set of trains, $\mathcal{S}_j$ the set of stations passed by $j$, $w_j$ is the particular weight reflecting the priority of the $j$'th train. For implementation reasons it is more convenient to use:
\begin{equation}\label{eq::fd}
    f = \sum_{j \in \mathcal{J}} w_j \sum_{s \in \mathcal{S}_j} \frac{d_a(j, s_{\text{out}})}{d_{\text{max}}}.
\end{equation}

For clarity of presentation, we introduce the minimal time train $j$ is ready to leave $s$ provided the initial conditions and that no other trains are on the route and denote it by $\upsilon(j,s_{\text{out}})$. By definition, 
\begin{equation}\label{eq::udefn}
\upsilon(j,s_{\text{out}}) =  \sigma (j, s_{\text{out}}) + d_u(j, s_{\text{out}})
\end{equation} 
and
\begin{align}\label{eq::t-d}
d_a(j, s_{\text{out}}) &= d(j, s_{\text{out}}) - d_u(j, s_{\text{out}})  \nonumber\\
&= t(j,s_{\text{out}}) - \sigma (j, s_{\text{out}}) - d_u(j, s_{\text{out}}) \nonumber \\
&= t(j,s_{\text{out}}) - \upsilon(j,s_{\text{out}}),
\end{align} 
where the first line follows by  Eq.~\eqref{eq::split_d}, the second line follows by Eq.~\eqref{eq::delay_repr} and the third line follows by Eq.~\eqref{eq::udefn}.

Now we can rewrite the objective function defined in Eq.~\eqref{eq::fd} using Eq.~\eqref{eq::t-d} as
\begin{equation}\label{eq::objective}
    f = \sum_{j \in \mathcal{J}} w_j \sum_{s \in \mathcal{S}_j} \frac{t(j,s_{\text{out}}) - \upsilon(j,s_{\text{out}}) }{d_{\text{max}}}.
\end{equation}
As the objective is defined, we move on to constraints derived from train traffic safety conditions and other technical issues.

\begin{figure}[tbh!]
    \centering
    \includegraphics[width = 0.8\linewidth]{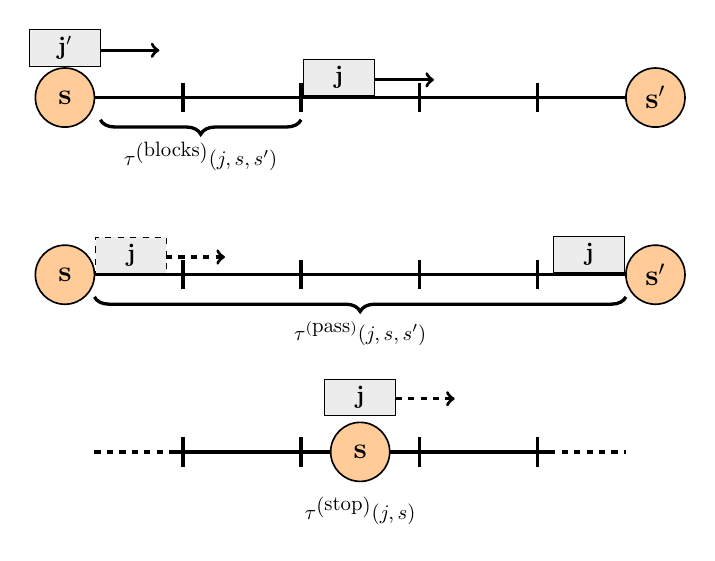}
    \caption{Illustration of $\tau^{(\textrm{blocks})}$, $\tau^{(\textrm{pass})}$ and $\tau^{(\textrm{stop})}$, in our model they are in time units. In this demonstrative example $\tau^{(\textrm{blocks})}$ requires passing two subsequent block sections, which is rather usual for trains traffic management, but not the limitation of the model. (We do not consider here the length of the train.)}
    \label{fig:my_label}
\end{figure}

We start with the \textbf{minimal passing time} condition which ensures that for any pair of subsequent stations $(s,s')$, that is on the route of $j \in J$, the entry time to station $s'$ is exactly equal to the leaving time of station $s$ plus the time it takes for train $j$ to move from $s$ to $s'$, which we denote by $\tau^{(\text{pass})}(j, s, s')$, see also Fig.~\ref{fig:my_label}. Note that we make an assumption that the train can leave $s$ only if it can proceed at full speed to $s'$. Given this, the condition can be stated as:
\begin{equation}\label{eq::min_pass_time}  
     t(j, s'_{\text{in}}) =  t(j, s_{\text{out}}) + \tau^{(\text{pass})}(j, s, s').
\end{equation}

Next, we move to the \textbf{minimal headway} condition. Consider trains $j,j'$ heading in the same direction. To determine their order, we use the  precedence variables $ y(j,j',s_{\text{out}}) \in \{0,1\} $ that is equal to 1 iff $j$ leaves $s$ before $j'$. (The precedence variable implementation appears to be more efficient than the order variable implementation~\cite{lange2018approaches}.) Naturally, for any $j,j' \in \mathcal{J}$ and $s \in \mathcal{S}_j \bigcap \mathcal{S}_{j'}$, it follows that
\begin{equation}\label{eq::y_one_way} 
    y(j,j',s_{\text{out}}) = 1 - y(j',j,s_{\text{out}}).
\end{equation}

Assume that train $j$ leaves $s$ before train $j'$. Then $j'$ needs to wait for at least additional $\tau^{(\text{blocks})}(j, s, s')$ which is the minimal time (headway) required for train $j$ (traveling from $s$ to $s'$) to release blocks to allow $j'$ to follow at full speed, see also simple illustrative presentation in Fig.~\ref{fig:my_label}. However, if $j$ is slower than $j'$, then an additional waiting time of $ \tau^{(\text{pass})}(j, s, s') - \tau^{(\text{pass})}(j', s, s')$ is needed. For all ${j,j' \in \mathcal{J}^d}$ -- the set of pairs of trains heading toward the same direction on the same route -- and ${ (s,s') \in \mathcal{C}_{j,j'}}$ -- the set of subsequent stations in the common route of $j$ and $j'$ -- the condition can be expressed as follows:

\begin{equation}
\begin{split}
&y(j,j',s_{\text{out}}) = 1 \implies  \\
&{t(j', s_{\text{out}}) \geq  t(j, s_{\text{out}}) + \tau^{(\text{blocks})}(j, s, s')+\max\{0,  \tau^{(\text{pass})}(j, s, s') - \tau^{(\text{pass})}(j', s, s') \}}.
\label{eq::spacing_at_line}
\end{split}
\end{equation}

In a single track line, a train can enter the single line only if it is cleared by the train approaching from the opposite direction -- we call it the \textbf{single track line condition}. Similar to $y$, we define the precedence variable 
$z(j,j',s,s') \in \{0,1\}$, that determines which train enters first the single track line between $s$ and $s'$. Note that the following is true for all $j,j'  \in \mathcal{J}^{o}_{\text{single}}$  -- the set of all trains heading in opposite direction on the same track -- and $(s,s') \in \mathcal{C}_{j,j'}$.
\begin{equation}\label{eq::y_two_ways}
z(j,j',s,s') + z(j',j,s',s) = 1.
\end{equation}
By $\tau^{(\text{res.})}(j, s)$, we denote the time of using the conflicted resource (i.e. set of switches) by trains $j$ at station $s$, see also Fig~\ref{fig:my_label}. For all $j,j' \in  \mathcal{J}^{o}_{\text{single}}$ and $ (s,s') \in \mathcal{C}_{j,j'}$, the \textbf{single track line} condition is expressed as:
\begin{equation}\label{eq::single_line}
    z(j,j',s,s') = 1 \implies t(j', s'_{\text{out}}) \geq t(j, s'_{\text{in}}) + \tau^{(\text{res.})}(j , s'),
\end{equation}

If the train is due to stop at the station,  then it needs to wait at least $\tau^{(\text{stop})}(j,s)$, which is the minimal stopping time at the station $s$ by train $j$, see Fig.~\ref{fig:my_label}. Apart from this, the train must not leave before its scheduled departure time. This is called the \textbf{minimal stay} condition. This results in the following conditions for all $j \in \mathcal{J}$ and $s \in S_j$: 
\begin{equation}\label{eq::enter_platforms}
 t(j,s_{\text{out}}) \geq t(j,s_{\text{in}}) + \tau^{(\text{stop})}(j, s),
\end{equation}
and
\begin{equation}\label{eq::departure_time}
    t(j,s_{\text{out}}) \geq \sigma(j,s_{\text{out}}).
\end{equation}

\begin{figure}[tbh!]
    \centering
    \includegraphics[width = 0.8\linewidth]{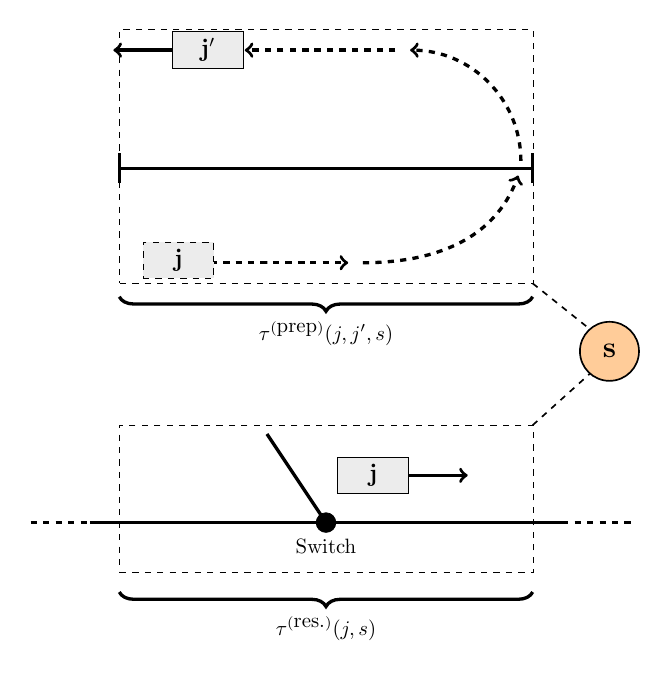}
    \caption{Illustration of $\tau^{(\textrm{res})}$ and $\tau^{(\textrm{prep})}$. Train $j$ terminates at station $s$ and the rolling stock is changed to another train $j'$ (upper panel). Train $j$ occupies switch at station $s$, and such switch is not available for other train at that time (lower panel).}
    \label{fig:my_label2}
\end{figure}

We also use the \textbf{rolling stock circulation} condition analogous to the one discussed in~\cite{domino2020quantum}. By $\tau^{\text{(prep.)}}(j,j', s)$ we denote the minimal rolling stock preparation time, if train $j$ terminates at $s$ and then starts as new $j'$, see Fig.~\ref{fig:my_label2}.
For all ${s \in S} $ and ${j,j' \in \mathcal{J}_s^{\text{round}}}$ -- the set of pairs of trains that terminates at $s$ and set off as a new train -- we have the condition:
\begin{equation}\label{eq::rolling_stock}
t(j', s_{\text{out}}) \geq t(j, s_{\text{in}}) + \tau^{\text{(prep.)}}(j,j', s).
\end{equation}

There are cases where two trains are to use the same set of switches at station $s$ while entering the station, and leaving it. This is called the \textbf{switch occupancy} condition. 
This condition is (partially) integrated with the single track line condition (a common set of switches where a single line enters/ leaves a station) and track occupancy condition (a common track that can be occupied by one train only). Hence as $\mathcal{J}_s^{\text{switch}}$ we consider the set of pairs of trains that compete for the same switch or switch set not considered in other conditions. For all $s \in S$ and $j,j' \in \mathcal{J}_s^{\text{switch}}$, this condition can be stated as:
\begin{equation}\label{eq::switch_cond}
     y(j,j', s^{***}) = 1 \implies t(j', s^*) \geq t(j, s^{**}) + \tau^{(\text{res.})}(j, s),
\end{equation}
where $s^{*}$, $s^{**}$, $s^{***}$ may be $s_{\text{in}}$ or $s_{\text{out}}$ depending on the particular situation on the station.
Two trains can not occupy \textbf{set of switches} at the station -- Eq.~\eqref{eq::switch_cond}
In Eq.~\eqref{eq::switch_cond}, $s^{*}$, $s^{**}$ may be $s_{\text{in}}$ or $s_{\text{out}}$ depending on the particular trains at a station, similarly $y(j', j, s^{***})$. For example if $j$ and $j'$ compete for the common switch as $j$ and $j'$ both leave $s$, we have $s^{*} = s^{**} = s^{***} = s_{\text{out}}$. There may be also other possibilities, e.g. including $z$ variable instead of $y$ variable, however we do not discuss them in this simple model. 

Now, let's discuss the \textbf{track occupancy} condition. As we are using a parallel machine approach, trains are assigned to particular tracks and station blocks that can be occupied only by one train at once. Consider two trains $j_1, \; j_2$ that compete for the same track at the station. The subsequent train has to wait until the previous one leaves. This results in
\begin{equation}\label{eq::platform2}
 y(j, j', s_{\text{out}}) = 1 \implies \\ 
t(j', s_{in}) \geq t(j, s_{out}) + \tau^{(\text{res.})}(j, s)
\end{equation}
for all ${s \in S}$ and $j, j' \in \mathcal{J}^{\text{track}}_s$. Here $\mathcal{J}^{\text{track}}_s$ is the set of trains that compete for the same track at station $s$.
The additional term $\tau^{(\text{res.})}$ can be used if the two above-mentioned trains use the same set of switches (then the pair is excluded from $\mathcal{J}_s^{\text{switch}}$).

\begin{table}[]
	\centering
	\begin{tabular}{lp{0.58\textwidth}}
		\textbf{Symbol} & \textbf{Description}  
		\\ \toprule
		$j\in \mathcal{J}, (j,j')$ & train, pair of trains \\ \hline
		$\mathcal{J}^d$, $\mathcal{J}^{o}_{\text{single}}$  & set of trains heading in the same direction on the same route, in opposite directions on the same track \\ \hline
	    $ \mathcal{J}^{\text{track}}_s, \mathcal{J}_s^{\text{switch}}$ &  set of trains that compete for the same station block, switch at station \\ \hline
	    $\mathcal{J}_{s}^{ \text{round}}$ & the set of all pairs of trains such that $j$ terminates at $s$ turn around and starts from $s$ as $j'$  \\ \hline
	    $s\in \mathcal{S}, (s, s')$ & station, pair of stations\\ \hline
	    $\mathcal{S}_j$ & set of all stations in the route of train $j$  \\ \hline
		$\mathcal{C}_j, \mathcal{C}_{j, j'}$ & set of all subsequent pairs of stations in the route of $j$, common route of $j$, $j'$\\ \hline
		$\sigma(j,s_{\text{out}})$  & scheduled time of entering, leaving station $s$ by train $j$\\ \hline
		$\upsilon(j,s_{\text{out}})$ & minimal time the train $j$ is ready to leave $s$, provided the initial conditions and that no other trains are on the route \\ \hline
		$d(j,s_{\text{out}})$, $d_u(j,s_{\text{out}})$, $d_a(j,s_{\text{out}})$ & delay, unavoidable delay, additional delay of train $j$ on leaving station $s$ \\ \hline
		$d_{\text{max}}$ & maximum possible (acceptable) additional delay  \\ \hline
		
		$N(d_{\max})$ & number of trains each train may be in conflict at each station, track or switch (on average) \\ \hline
		$\tau^{(\text{pass})}(j, s, s')$ & minimal passing time of train $j$ between $s$ and $s'$ (the time it takes train $j$ to travel from $s$ to $s'$) \\ \hline
        $\tau^{(\text{blocks})}(j, s, s')$ & minimal time required for train $j$ (traveling from $s$ to $s'$) to release blocks to allow another train to follow at a top speed \\ \hline
        $\tau^{(\text{stop})}(j,s)$ & minimal stopping time at the station $s$ by train $j$ \\ \hline
        $\tau^{\text{(prep.)}}(j,j', s)$ &minimal rolling stock preparation time \\ \hline     $\tau^{(\text{res.})}(j, s)$ & time of using the conflicted resource (i.e. set of swishes) by trains $j$ at stations $s$ \\
        \hline     $w_j$ & weight of train $j$ in the objective \\
        \hline     $p_{\text{sum}}$, $p_{\text{pair}}$, $p_{\text{qubic}}$  & penalty constants for HOBO / QUBO formulation. \\
        \hline     $f$ & objective function. \\
        \hline     $M$ & a large constant for linearization. \\
        \bottomrule
	\end{tabular}
	\caption{Summary of the notations used in the paper to denote parameters of the model.}\label{tab::symbols}
\end{table}

\begin{table}[]
	\centering
	\begin{tabular}{llp{0.7\textwidth}}
		\textbf{Symbol} &  \textbf{Type} & \textbf{Description}  
		\\ \toprule
		$t(j,s_{\text{out}})$& integer &  time of train $j$ on leaving  station $s$ \\ \hline
         $t(j,s_{\text{in}})$ & integer &  time of train $j$ on entering station $s$, uniquely determined by $t(j,s_{\text{out}})$ \\ \hline
		$y(j, j,s_{\text{out}})$ & binary 0-1 & 1 iff $j$ leaves $s$ before $j'$ (determines the order of trains $j$ and $j'$ while leaving station $s$) \\ \hline
		$z(j, j',s, s')$ & binary 0-1 & 1 iff train $j$ enters the single track line between $s$ and $s'$ before $j'$. (determines the order of trains $j$ and $j'$ while entering the particular track  line between station $s$ and $s'$) \\ \hline
		$x_{j,t,s}$ & binary 0-1 & $1$ iff train $j$ leaves station $s$ at time $t$. \\ 
		\hline
		$\tilde{x}_{j, j', t, t', s}$ & binary 0-1 & auxiliary variable for HOBO quadratisation $\tilde{x}_{j, j', t, t', s} = x_{j,t,s}  x_{j',t',s}$ \\ 
        \bottomrule
	\end{tabular}
	\caption{List of the variables used in the paper.}\label{tab::var_symbols}
\end{table}

\subsection{Integer linear programming representation}

Based on the problem formulation presented above, we construct an integer linear programming (ILP) formulation. To linearize the implications, of the form $a=1 \implies b \geq c$, we use the transformation $b + M(1-a) \geq c$ where $M$ is a large constant. Furthermore, we use Eq.~\eqref{eq::y_one_way} and Eq.~\eqref{eq::y_two_ways} for the simplification of the equations with precedence variables. We use the variables $t(j,s_{\text{out}})$, $y(j,j',s_{\text{out}})$ and $z(j,j',s,s')$ as defined previously. ILP takes the following form.

\begin{flalign}
&\textrm{min.}   ~\sum_{j \in \mathcal{J}} w_j \sum_{s \in S_j} \frac{t(j,s_{\text{out}}) -\upsilon(j,s_{\text{out}}) }{d_{\text{max}}} \label{eq::obj} \\
&\textrm{subject to}  \nonumber \\ 
&~~~   t(j, s'_{\text{in}}) -  t(j, s_{\text{out}}) = \tau^{(\text{pass})}(j, s, s') 
&\forall_{j \in \mathcal{J}} \  \forall_{(s, s') \in \mathcal{C}_j}  \label{eq:c1}  \\
&~~~  t(j', s_{\text{out}}) + M \cdot (1-y(j,j',s_{\text{out}})) - t(j, s_{\text{out}}) & \nonumber \\ 
\rlap{~~~~~~~~$\geq  \tau^{(\text{blocks})}(j, s, s') 
    +   \max\{0,  \tau^{(\text{pass})}(j, s, s') - \tau^{(\text{pass})}(j', s, s') \} $}  \nonumber \\ 
&&\forall_{j,j' \in \mathcal{J}^d}  \ \forall_{(s,s') \in \mathcal{C}_{j,j'}} \label{eq:c2}\\
&~~~  t(j', s'_{\text{out}}) + M\cdot z(j,j',s, s')- t(j, s'_{\text{in}}) \geq \tau^{(\text{res.})}(j, s')  &\nonumber \\
&& \forall_{j,j' \in \mathcal{J}^{o}_{\text{single}}} \forall_{(s,s') \in \mathcal{C}_{j,j'}} \label{eq:c3} \\
&~~~t(j,s_{\text{out}}) - t(j,s_{\text{in}})  \geq  \tau^{(\text{stop})}(j, s) &\forall_{j \in \mathcal{J}} \ \forall_{s \in S_j} \label{eq:c4}\\
&~~~    t(j,s_{\text{out}}) \geq \sigma(j,s_{\text{out}}) &\forall_{j \in \mathcal{J}} \ \forall_{s \in S_j} \label{eq:c5} \\
&~~~t(j', s_{\text{out}}) - t(j, s_{\text{in}}) \geq  \tau^{\text{(prep.)}}(j,j', s) & \forall_{s \in S} \forall_{(j,j') \in \mathcal{J}_s^{\text{round}}} \label{eq:c6}\\
&~~~ t(j', s_{\text{in}}) + M \cdot y(j,j',s_{\text{out}}) - t(j, s_{\text{out}})  \geq \tau^{(\text{res.})}(j, s) & \forall_{s \in S} \forall_{j,j' \in \mathcal{J}^{\text{track}}_{s}} \label{eq:c7} \\
&~~~ t(j', s^{*}) + M \cdot y(j,j',s^{***}) - t(j, s^{**})  \geq \tau^{(\text{res.})}(j, s) & \forall_{s \in S} \forall_{j,j' \in \mathcal{J}_s^{\text{switch}}} \label{eq:c8} \\
  \nonumber \\
&~~~y(j',j,s_{\text{out}}) \in \{0,1\},~ z(j,j',s,s') \in \{0,1\}  &  \forall_{(s,s') \in \mathcal{C}_{j,j'}} \\ 
&~~~t(j,s_{\text{out}}) \in \{\upsilon(j,s_{\text{out}}), \dots, \upsilon(j,s_{\text{out}}) + d_{\text{max}}\} & \forall_{j \in \mathcal{J}} \ \forall_{s \in S_j}
\end{flalign}

The range for the integer variables $t(j,s_{\text{out}})$ follows since the following is true by Eq.~\eqref{eq::delay_repr}, Eq.~\eqref{eq::d_limit}, and the definition of $\upsilon(j,s_{\text{out}})$.

\begin{align}
 t(j,s_{\text{out}}) &= d_u(j,s_{\text{out}}) + d_a(j,s_{\text{out}})  + \sigma(j,m) = \upsilon(j,s_{\text{out}}) +
 d_a(j,s_{\text{out}})  \label{eq::trange1} \\
 \upsilon(j,s_{\text{out}})&\leq \upsilon(j,s_{\text{out}}) +
 d_a(j,s_{\text{out}}) \leq   \upsilon(j,s_{\text{out}}) + d_{\text{max}} \label{eq::trange2}
\end{align}

Although we use the variables $t(j,s_{\text{in}})$ for the clarity of the formulation, thanks to the first constraint, they are defined uniquely and not used when formulating the program. Given this, we have roughly a single time variable per station and train (but some trains may not serve all stations) and overall : 
 \begin{equation}
     \#(t) \leq \vert\mathcal{J}\vert \vert\mathcal{S}\vert.
     \label{eq::no_t}
 \end{equation}
 
Similarly, we define the precedence variables only for an ordered pair $(j,j')$ as the corresponding variable can be replaced using Eq.~\eqref{eq::y_one_way} and Eq.~\eqref{eq::y_two_ways}. This results in a single precedence variable $y$ per station and train pair.
\begin{equation}\label{eq::jlinear}
  \#(y) \leq \frac{\vert \mathcal{J}\vert \left(\vert \mathcal{J}\vert  - 1 \right)}{2} \vert \mathcal{S}\vert .
\end{equation}
However, we do not need to compare all pairs of trains in case of dense train traffic, and the number of trains to be compared is somehow limited by $d_{\max}$. (There will be pairs that would never meet for given $d_{\max}$). Let assume each train can be in conflict with $N(d_{\max}) \leq \frac{\vert \mathcal{J}\vert -1}{2}$ trains at each station, track or switch (on average). $N(d_{\max})$ is non-decreasing in $d_{\max}$. We have then the approximation:
\begin{equation}\label{eq::jlinear_approx}
  \#(y) \approx \vert \mathcal{J}\vert  \vert \mathcal{S}\vert  N(d_{\max})
\end{equation}
We also have some additional precedence variables e.g. for the single-track lines. Using similar approximation: \begin{equation}\label{eq::nz_approx}
  \#(z) \approx \vert \mathcal{J}\vert  \vert \mathcal{S}\vert  N(d_{\max})
\end{equation}
This is however adequate if all trains use the single track line, otherwise, we can treat it as the limit. 

The number of \textbf{minimal headway} Eq.~\eqref{eq:c2}, and \textbf{track occupancy} Eq.~\eqref{eq:c7} constraints are both roughly equal to number of $y$ variables, as each such variable concerns the conflict on these conditions. The number of \textbf{single track line} Eq.~\eqref{eq:c3} conditions is roughly proportional to number of $z$ variables from the same reason. The number of \textbf{minimal stay} constraints Eq.~\eqref{eq:c4} and Eq.~\eqref{eq:c5} are both limited (or can be approximated) by $\vert \mathcal{S}\vert $ $\vert \mathcal{J}\vert $ (limit comes from the fact that not all trains serve all stations).

The number of \textbf{rolling stock circulation} constraints Eq.~\eqref{eq:c6} is not large in comparison with others, for sure it is limited by $\frac{\vert \mathcal{J}\vert }{2}$ (this would be a situation that one-half of the trains turn to another half). The number of variables in \textbf{switch conditions}, Eq.~\eqref{eq:c8} is not straightforward, as there are many possibilities and approaches. We can again approximate them by the number of $y$ variables. The number of constraints can be approximated/ limited by:
\begin{equation}
\label{eq:: ilp_ilpterms}
\begin{split}
    \#(\text{constraints}) &\approx 3 \#(y) + \#(z) + 2 \vert \mathcal{S}\vert  \vert \mathcal{J}\vert  \\ &\approx (4 N(d_{\max}) + 2) \vert \mathcal{S}\vert  \vert \mathcal{J}\vert  \geq \left(4 \frac{\vert \mathcal{J}\vert -1}{2} + 2\right) \vert \mathcal{S}\vert  \vert \mathcal{J}\vert .
    \end{split}
\end{equation}
Hence, one can conclude that if $d_{\max}$ is set properly, the problem size should be linear in the number of trains and stations.

It is broadly accepted that railway problems are equivalent to job-shop models with blocking constraints, see eg.~\cite{szpigel1973optimal} (such job-shop is equivalent in principle to the set partition problem - see eg.~\cite{sotskov1995np}). In detail, in such an NP-hard problem, we have the release $t_i$ and due dates $\upsilon_i$ of jobs, requirements of the model (blocking constraints), and there may be also some additional constraints such as no-waiting, and recirculation (rcrc). In our analogy, trains are jobs, and selected block sections are machines. With the standard notation of scheduling
theory~\cite{pinedo2012scheduling}, our problem falls into the class $\mathcal{J}\vert t_i, \upsilon_i, $block, no wait, rcrc$\vert  \sum_j w_j T_j$. Above mentioned conditions comply with ours in the following way:
\begin{enumerate}
    \item Eq.~(\ref{eq::obj}) is the objective, weighted tardiness with incorporated due time $\upsilon$, 
    \item Eq.~(\ref{eq:c1}) is the no-waiting constraint on the line (on the station waiting is allowed),
    \item Eqs.~(\ref{eq:c2})~(\ref{eq:c3})~(\ref{eq:c7})~(\ref{eq:c8})  are blocking constraints,
    \item Eq.~\eqref{eq:c6} is the recirculation constraint;
    \item Eqs.~\eqref{eq:c4}~\eqref{eq:c5} concern the release time.
\end{enumerate}

The presented linear programming approach is a standalone model. However, it fails in rapid computation for some models with more than a few trains~\cite{lange2018approaches}. Hence it may be beneficial to use another computation paradigm, such as quantum (or quantum-inspired) annealing.
As the alternative, in the
next subsection we derive the HOBO representation directly form dispatching
conditions (i.e. independently on ILP).

\subsection{HOBO representation}

A higher-order unconstrained binary optimization (HOBO) problem involves the minimization of a multilinear polynomial expression defined over binary variables
\begin{equation*}
h(\mathbf{x}) = \sum_{S \subseteq V} c_S \prod_{i\in S} x_i, 
\end{equation*}
where $\mathbf{x}$ denotes the vector of all binary variables $x_1,x_2,\dots,x_n$, $V=\{1,2,\dots,n \}$ and $c_S$ are the real coefficients. It is also equivalently expressed as Pseudo-Boolean optimization \cite{boros2002pseudo} and polynomial unconstrained binary optimization \cite{glover2011polynomial}. 

The degree or order of a HOBO is the size of the largest set $S$. The problem is called quadratic unconstrained binary optimization (QUBO) when the degree is equal to 2, and the term HOBO is often used for higher-order problems. For the parallel machine approach adopted in this paper, we have the third order of HOBO. 

To formulate the problem, we use the time indexing variable
\begin{equation}
x_{j,t,s} \in \{0,1\},
\end{equation}
that is 1 if train $j$ leaves station $s$ at time $t$, and $0$ otherwise (recall that each time index $t$ can be represented uniquely by delay via Eq.~\eqref{eq::delay_repr}). Note that, we use Eq.~\eqref{eq::min_pass_time} to compute the arrival time from the departure time from the previous station. We use the discretised $t$ that is limited from both sides by Eq.~\eqref{eq::trange1} and Eq.~\eqref{eq::trange2}.  We denote this limit by 
\begin{equation}
t \in \mathcal{T}_{j,s}, \text{ where } \mathcal{T}_{j,s} \equiv \{\upsilon(j,s_{\text{out}}),  \upsilon(j,s_{\text{out}})+1, \ldots,  \upsilon(j,s_{\text{out}}) + d_{\max}\},
\end{equation}
here we consider one-minute resolution. This limitation ensures the timetable condition in Eq.~\eqref{eq::departure_time}.

We have the linear objective function defined as in Eq.~\eqref{eq::objective}:
\begin{equation}
    f(\mathbf{x}) = \sum_{j \in \mathcal{J}} w_j \sum_{s \in S_j} \sum_{t \in \mathcal{T}_{j,s}} \frac{t(j,s_{\text{out}}) - \upsilon(j,s_{\text{out}}) }{d_{\text{max}}} x_{j, t, s}.
\end{equation}

In our approach, we do not take into account recirculation, i.e. each train leaves each station $s \in S_j$ once and only once:
\begin{equation}\label{eq::q_linear}
 \forall_{j \in \mathcal{J}} \forall_{s \in S_j} \sum_{t}  x_{j,t,s} = 1.
\end{equation}

To convert the constrained problem into an unconstrained one, we use the well-established penalty method~\cite{luenberger2015linear}. Constraints are incorporated into the objective function so that violation of the constraints adds a positive penalty to the objective function. For instance, to include the constraint in Eq.~\eqref{eq::q_linear} in the objective function, we set a large enough penalty constant $p_{\text{sum}}$ and use the following penalty term:
\begin{equation}\label{eq::linear}
    P_{\text{sum}}(\mathbf{x}) = p_{\text{sum}} \sum_{j \in \mathcal{J}, s \in S_j} \left( \sum_{t,t' \in \mathcal{T}_{j,s}^{\times 2} \ t \neq t'} x_{j,t,s} x_{j,t',s}  - \sum_{t \in \mathcal{T}_{j,s}} x_{j,t,s} \right).
\end{equation}
Following \cite{domino2020quantum}, the  conditions described in Eq.~\eqref{eq::min_pass_time} -- \eqref{eq::platform2} can be expressed using binary variables so that the quadratic terms yield $0$ if the solution is feasible, and produces a penalty otherwise. For this reason, we use a sufficiently large penalty constant $p_{\text{pair}}$. Note that we have symmetric terms  ($x_1 x_2 + x_2 x_1$) to follow the convention of symmetric QUBO formulation.

The \textbf{minimal headway} condition given by  Eq.~\eqref{eq::y_one_way} and Eq.~\eqref{eq::spacing_at_line},  can be expressed in the following form:

\begin{flalign}
 &P_{\text{pair}}^{\text{headway}}(\mathbf{x}) =  p_{\text{pair}} \sum_{j,j' \in \mathcal{J}^d}  \ \sum_{(s,s') \in \mathcal{C}_{j,j'}} \
   \sum_{t \in \mathcal{T}_{j,s}, t' \in \mathcal{T}_{j', s}, A < t' - t < B} (x_{j,t,s} x_{j', t', s} + x_{j',t',s} x_{j, t, s}), \nonumber \\
&\noindent \text{ where }& \nonumber \\
&\rlap{\hspace{0.6in}$A = -\tau^{(\text{blocks})}(j', s, s') - \max\{0,  \tau^{(\text{pass})}(j', s, s') - \tau^{(\text{pass})}(j, s, s') \},$} \nonumber\\
&\rlap{\hspace{0.6in}$B =  \tau^{(\text{blocks})}(j, s, s') + \max\{0,  \tau^{(\text{pass})}(j, s, s') - \tau^{(\text{pass})}(j', s, s') \}.$}\nonumber \\
\end{flalign}

The \textbf{single track} condition defined in Eq.~\eqref{eq::y_two_ways} and Eq.~\eqref{eq::single_line} yields:
\begin{flalign}
   &P_{\text{pair}}^{1\text{track}}(\mathbf{x}) =
  p_{\text{pair}} \sum_{j,j' \in \mathcal{J}^{o}_{\text{single}}} \sum_{(s,s') \in \mathcal{C}_{j,j'}} \sum_{\substack{t \in \mathcal{T}_{j,s}, t' \in \mathcal{T}_{j', s'} \\A < t' - t <  B}} (x_{j, t, s} x_{j', t', s'} + x_{j', t', s'} x_{j, t, s}), \nonumber \\
   &\noindent \text{ where } & \nonumber \\
   &\rlap{\hspace{0.7in}$A = - \tau^{(\text{res})}(j,j',s')  - \tau^{(\text{pass})}(j',s',s'),$} \nonumber \\
   &\rlap{\hspace{0.7in}$B = \tau^{(\text{pass})}(j,s,s')  + \tau^{(\text{res})}(j,j',s').$} \nonumber \\ \label{eq::single_line_qubo}
\end{flalign}
   
The \textbf{minimal stay} condition given in Eq.~\eqref{eq::enter_platforms} (incorporated if necessary with Eq.~\eqref{eq::departure_time}) yields:
\begin{equation}
P_{\text{pair}}^{\text{stay}}(\mathbf{x}) = p_{\text{pair}} \sum_{j \in \mathcal{J}}  \sum_{(s,s') \in \mathcal{C}_j} 
    \sum_{\substack{t \in \mathcal{T}_{j,s}, t' \in \mathcal{T}_{j, s'} \\ t' < t + \tau^{(\text{pass})}(j,s,s') + \tau^{(\text{stop})}(j,s)}} ( x_{j,t,s} x_{j, t', s'} + x_{j,t',s'} x_{j, t, s} ).
\end{equation}
The \textbf{rolling stock circulation} condition in Eq.~\eqref{eq::rolling_stock} yields:
\begin{flalign}
     &P_{\text{pair}}^{\text{circ}}(\mathbf{x}) =  
     p_{\text{pair}} \sum_{s \in S} \sum_{(j,j') \in \mathcal{J}_s^{\text{round}}}    \sum_{\substack{t \in \mathcal{T}_{j,s}, t' \in \mathcal{T}_{j, s'} \\ t' < t + \tau^{(\text{pass})}(j,s,s') + \tau^{(\text{prep.})}(j, j',s)}} ( x_{j,t,s} x_{j, t', s'} + x_{j,t',s'} x_{j, t, s} ).
\end{flalign}

The \textbf{switch occupation} condition in Eq.~\eqref{eq::switch_cond} yields:

\begin{flalign}
&P_{\text{pair}}^{\text{switch}}(\mathbf{x}) = p_{\text{pair}} \sum_{s \in S} \sum_{j,j' \in \mathcal{J}_s^{\text{switch}}} \sum_{\substack{t \in \mathcal{T}_{j,s}, t' \in \mathcal{T}_{j', s}  \nonumber\\ -\tau^{(\text{res.})}(j',s) < t'-t < \tau^{(\text{res.})}(j,s)}} (x_{j,t,s} x_{j', t', s} + x_{j',t',s} x_{j, t, s}).\\
\label{eq::switch}
\end{flalign}

The above can be checked alone or integrated with other conditions such as \textbf{track occupation condition} 
in Eq.~\eqref{eq::platform2} and \textbf{single track} condition 
in Eq.~\eqref{eq::single_line_qubo}. The order of trains can be changed at the station only if these trains use different tracks at the station. Suppose that $j$ and $j'$ are on the same track at the station, hence they can not change order. To express this condition we need a higher order term, which yields a HOBO formulation. Let $t'' = t(j', s'_{\text{out}})$, $t' = t(j', s_{\text{out}})$ and $t = t(j, s_{\text{out}})$, where $s'$ is a station prior to $s$ in the route of $j'$. If $j$ leaves before $j'$, i.e. $t < t'$ ( $t \neq t'$ to prevent trains leaving the same track at the same time), then $j'$ must enter after $j$ leaves i.e. $t'' +  \tau^{(\text{pass})}(j', s', s)  \geq t + \tau^{(\text{res})}(j, j', s) $. The following term needs to be 0:
\begin{equation}\label{eq::p_qubic}
P_{\text{qubic}}^{\text{occ.}}(\mathbf{x}) = 2p_{\text{pair}} \sum_{s \in S} \sum_{j, j' \in \mathcal{J}^{\text{track}}_s}     \sum_{\substack{t \in \mathcal{T}_{j,s}, \ t' \in \mathcal{T}_{j', s} \ t'' \in \mathcal{T}_{j', s'}  \\ t'' +  \tau^{(\text{pass})}(j', s', s)  - \tau^{(\text{res})}(j, j', s) < t \leq t' }}  x_{j,t,s}  x_{j',t',s}  x_{j', t'', s'}.
\end{equation}
We use the penalty value $2p_{\text{pair}}$ to be consistent with the symmetrization. 

The resulting HOBO representation is expressed as:
\begin{align}
    \text{min.}~~h(\mathbf{x}) = &f(\mathbf{x}) + P_{\text{sum}}(\mathbf{x}) + P_{\text{pair}}^{\text{headway}}(\mathbf{x}) + P_{\text{pair}}^{1\text{track}}(\mathbf{x}) + P_{\text{pair}}^{\text{stay}}(\mathbf{x}) +P_{\text{pair}}^{\text{circ}}(\mathbf{x}) \nonumber \\ 
    & + P_{\text{pair}}^{\text{switch}}(\mathbf{x}) +  P_{\text{qubic}}^{\text{occ.}}(\mathbf{x}), \nonumber \\
\end{align}
where $f(\mathbf{x})$ is the objective function and the rest are the penalty terms that need to be minimized. 

The penalty constants $p_{\text{sum}}$ and $p_{\text{pair}}$ has be large enough to ensure the constraints to be always fulfilled, regardless the penalty value in the objective. However, these constants cannot be too high; because, in that case, they may affect the performance of the quantum annealer.

The number of variables $x_{j,t,s}$ depends on the time resolution of the system and $d_{\max}$. It can be approximated by:
 \begin{equation}
     \#(x) \leq \vert \mathcal{J}\vert   \vert \mathcal{S}\vert  (d_{\max} + 1).
 \end{equation}
Here ``$\leq$" sign is used as some trains may not serve some stations. The number of variables here (as in ILP case Eq.~\eqref{eq::no_t}) should be linear in the number of trains and stations, but it is also proportional to $(d_{\max} + 1)$. This can be however contrasted with the fact that in the ILP case, we have a broader range of time-indexed variables.
As argued before, we assume that each train can be potentially in conflict with $N(d_{\max})$ at each station, line or switch. Then for each condition there are roughly $\vert \mathcal{J}\vert   \vert \mathcal{S}\vert  (d_{\max} + 1) N(d_{\max})$ non-zero QUBO or HOBO terms.
The exception is the \textbf{rolling stock occupation} condition, that occurs rather rarely in comparison with others. We expect roughly $3 \vert \mathcal{J}\vert   \vert \mathcal{S}\vert  (d_{\max} + 1) N(d_{\max})$ non-zero QUBO terms, and $ \vert \mathcal{J}\vert   \vert \mathcal{S}\vert  (d_{\max} + 1) N(d_{\max})$ HOBO terms. In comparison with LIP~(\ref{eq:: ilp_ilpterms}), this still can be an efficient implantation provided $d_{\max}$ is controlled.

\subsection{QUBO representation} 

A quadratic unconstrained binary optimization (QUBO) problem is formally defined as
 \begin{align*}\label{eq:qubo}
q(\mathbf{x}) =  \sum_{i,j = 1}^n x_i Q_{ij} x_j,
\end{align*}
where $Q$ is a real matrix of coefficients.
To be able to solve a problem using quantum annealing, we must first encode it using QUBO formulation as current quantum annealers allow only two-body interactions and representation through Ising model.

In this section, we will convert the HOBO representation into a QUBO representation. Note that we formulate HOBO directly from the dispatching conditions. The advantage of such a take is that in HOBO, we have one-to-one relation between real dispatching constraints and penalties of the mathematical formulation of the problem. (Latter auxiliary variables are only used in quadratization of HOBO). Alternatively, to obtain a QUBO formulation for the problem, one can transform the ILP presented in Section 3.1 by first converting inequalities into equalities using slack variables and then moving equality constraints to the objective using the penalty method. The ILP formulation requires binary variables quadratic in the number of trains. Furthermore, for the transformation, additional slack variables are needed as many as the number of inequality constraints which is quadratic in the number of trains, and they need to be optimized within the model as well. Since our HOBO approach is linear in the number of trains, we think that using it as the basis of the QUBO formulation may be more adequate for dense railroad traffic, with rather small delays; for instance for metro, trams, and urban rapid transport.

The qubic terms in the HOBO representation need to be converted to obtain a QUBO representation. The cubic terms can be expressed using quadratic terms at the cost of introducing new binary variables, see~\cite{mandal2020compressed}. For the decomposition, we use the auxiliary variable $ \tilde{x}_{j, j', t, t', s} = x_{j,t,s}  x_{j',t',s}$. The simplest approach here is to use the Rosenberg polynomial approach \cite{rosenberg1975reduction}. The constraint:
\begin{equation}
    x_{i_1} x_{i_2} x_{i_3} = 0,
\end{equation}
is equivalent to:
\begin{equation}\label{eq::z}
   \tilde{x}_{k} x_{i_3}  = 0, \text{ where } \tilde{x}_{k} = x_{i_1} x_{i_2},
\end{equation}
i.e. $k = k(i_1, i_2)$. Then one can use the polynomial:
\begin{equation}\label{eq::rosenberg_polynomial}
h(x_{i_1}, x_{i_2}, \tilde{x}_{k}) = 3 \tilde{x}_{k}^2 + x_{i_1} x_{i_2} -2 x_{i_1} \tilde{x}_{k} - 2 x_{i_2} \tilde{x}_{k},
\end{equation}
that is 0 if $ \tilde{x}_{k} = x_{i_1} x_{i_2}$, and positive (equal to $1$ or $3$) otherwise.
Using the auxiliary vector of variables $\mathbf{\tilde{x}}$, the penalty terms will be as follows:

\begin{flalign}
    &\rlap{$P_{\text{qubic}}(\mathbf{x},  \mathbf{\tilde{x}} ) = $} & \nonumber \\ &&p_{\text{pair}} \sum_{(i_1, i_2, i_3) \in \Gamma} \large( \tilde{x}_{k(i_1, i_2)} x_{i_3} + x_{i_3} \tilde{x}_{k(i_1, i_2)}  \large) + p_{\text{qubic}} \sum_{(i_1, i_2) \in \Gamma'} h(x_{i_1}, x_{i_2}, \tilde{x}_{k(i_1, i_2)}), \nonumber \\
\label{eg::p_qubic}
\end{flalign}

where $\Gamma$ is a set of particular indices of the cubic term (in Eq.~\eqref{eq::p_qubic}), and $\Gamma'$ a set of indices, where we require Eq.~\eqref{eq::z} to hold. Observe that for each pair of trains and for each station where the \textbf{track occupation condition} is to be checked; we have roughly $(d_{\text{max}}+1)^2$ auxiliary variables. Hence, this condition needs to be used with caution while modeling railway systems of considerable size.

The resulting QUBO representation is expressed as:
\begin{align}
    \text{min.}~~q(\mathbf{x},  \mathbf{\tilde{x}}) = &f(\mathbf{x}) + P_{\text{sum}}(\mathbf{x}) + P_{\text{pair}}^{\text{headway}}(\mathbf{x}) + P_{\text{pair}}^{1\text{track}}(\mathbf{x})  +P_{\text{pair}}^{\text{stay}}(\mathbf{x}) \nonumber \\ 
    & +P_{\text{pair}}^{\text{circ}}(\mathbf{x}) +P_{\text{pair}}^{\text{switch}}(\mathbf{x}) +  P_{\text{qubic}}(\mathbf{x},  \mathbf{\tilde{x}} ),\nonumber \\
\end{align}
where $f(\mathbf{x})$ is the objective function and the rest are the penalty terms that need to be minimized.

The number of $\tilde{x}$ variables $x_{j,t,s}$ depends on $d_{\max}$. It can be approximated by:
 \begin{equation}
     \#(\tilde{x}) \approx \vert \mathcal{J}\vert  \vert \mathcal{S}\vert  N(d_{\max}) (d_{\max} + 1)^2 ,
 \end{equation}
 as we use the same approximation as in~\eqref{eq::jlinear_approx}. We also have in mind that some trains may not serve some stations. When compared with Eq.~\eqref{eq::jlinear}, we can conclude that for the QUBO approach we need to control $d_{\max}$ more strictly. QUBO implementation may still be efficient but for small $d_{\max}$ determined, e.g., by some simple heuristics.
 
\subsection{Rerouting formulation}\label{s::Algorithm}

\begin{figure}[h]
\centering
\includegraphics[width=0.95\textwidth]{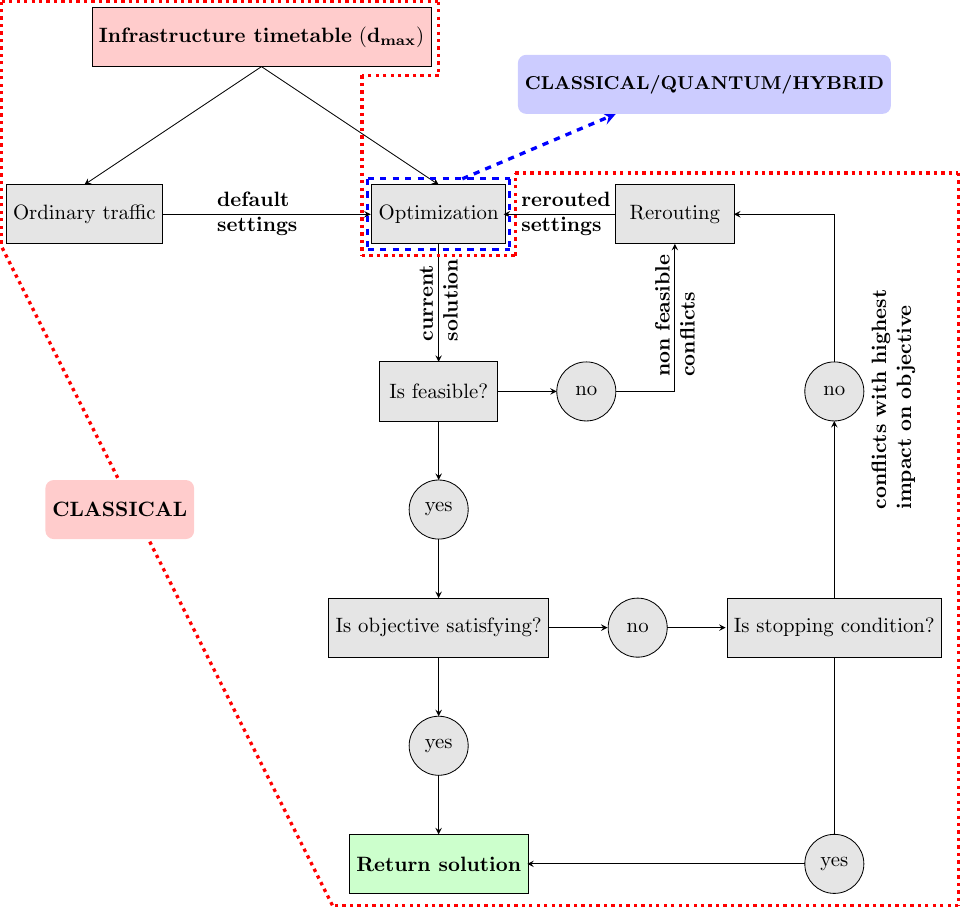}
\caption{The algorithm of quantum, classical, and hybrid quantum-classical for railway rescheduling and rerouting}
\label{fig::algorithm}
\end{figure}

We aim to solve the problem of setting the order of already delayed trains having limited resources in terms of infrastructure and traffic regulations. We follow the general idea set out in~\cite{shimada2021decomposition} where the widespread optimization problem needs to be decomposed into smaller components to demonstrate the supremacy of quantum or quantum-inspired approaches. In our case, we propose a decomposition that mimics some real-life rerouting practices. Namely, trains follow their default route as long as it does not cause distortion.  
Here we have the subproblem to be optimized by classical, quantum, or hybrid quantum-classical resources. If a solution is not satisfactory, we can change the path of the selected trains (aka reroute them) using the classical approach and then solve the new subproblem.
We propose the following algorithm summarized in Fig.~\ref{fig::algorithm}. The red region indicates the part that can be performed using the quantum (or quantum-inspired) resource at the current state of the art. As quantum computing becomes more and more advanced in the future, we will be moving the quantum border wider and wider to cover the whole algorithm finally.

We start from the given infrastructure, schedule, maximal possible additional delay parameters, priorities of the individual trains,  and the default train routes (aka default setting). Then we perform the optimization and check both feasibility of the solution as well as the objective value. If the solution is infeasible, we pick the nonfeasible conflict. Similarly, if we find the objective value too high, we pick the conflict, increasing the objective value the most. From this conflict, we pick one train (the one with lower priority) and reroute it by:
\begin{enumerate}
   \item changing the track to the parallel one,
    \item changing the platform at the station,
    \item changing the path within the station.
\end{enumerate}
We repeat the procedure until we get a satisfactory objective value or we achieve some stopping condition. The optimization subproblem (red) can be encoded either as a linear program, or following the QUBO or HOBO approaches.
 
\section{Demonstration of the model}\label{s::Demonstration}

\begin{figure}[h]
\centering
\includegraphics[width=0.63\textheight]{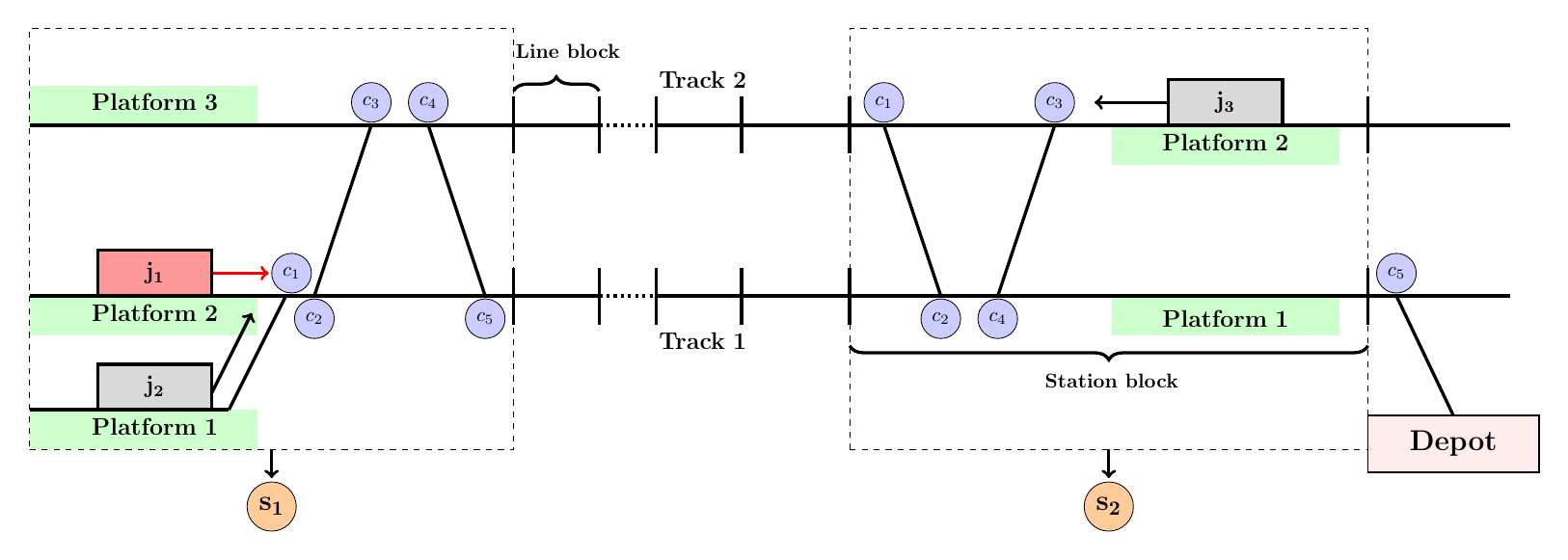}
\caption{The demonstrative model.}
\label{fig::toy1}
\end{figure}

We consider a railway model which we depict it in Fig.~\ref{fig::toy1}. There are $2$ stations $s_1$ and $s_2$, a double-track line  between them, and a depot; the switched are represented with $c_i$. We have $3$ trains: 
\begin{itemize}
    \item Inter-City (the faster one): $j_1$, $s_1 \rightarrow s_2$,
    \item Regional: $j_2$, $s_1 \rightarrow s_2$,
    \item Regional: $j_3$, $s_2 \rightarrow s_1$.
\end{itemize}

We will use the following parameters:
\begin{enumerate}
\item Minimal passing time: $\tau^{(\text{pass})}(j_1, s_1, s_2) = 4$, $\tau^{(\text{pass})}(j_2, s_1, s_2) = 8$, and $\tau^{(\text{pass})}(j_3, s_2, s_1) = 8$.
\item Minimal headway: $\tau^{(\text{blocks})}(j_1, s_1, s_2) = 2$ and $\tau^{(\text{blocks})}(j_2, s_1, s_2) = 2$.
\item Minimal stay: $\tau^{(\text{stop})}(j_1, s_2) = \tau^{(\text{stop})}(j_2, s_2) = 1$.
\item For all common resources, $\tau^{(res.)} = 1$.
\item After entering $s_2$, both $j_1$ and $j_2$ departs to the depot after the minimal stay. We only count delays of $j_1$ and $j_2$ at $s_1$ and delay of $j_3$ at $s_2$.
\end{enumerate}

Assume all trains are already delayed.  Hence, they can leave the stations as soon as the resources (free rail track ahead) are available. We consider the objective as denoted in Eq.~\eqref{eq::objective}, with the following weights $w_{j_1} = 2.0, w_{j_2} = w_{j_3} = 1.0$ (Inter-City train has higher priority). We set $d_{\max} = 10$ for all trains, and use $1$ minute resolution.
The initial conditions are as follows: $\nu(j_1, {s_1}_{\text{out}}) = 4$, $\nu(j_2, {s_1}_{\text{out}})=1$, and $\nu(j_3, {s_2}_{\text{out}}) = 8$. We compute unavoidable delays and $\nu$s prior to the optimization. Particular departure times of the trains are in the following range: 
\begin{equation}\label{eq::t_limits}
\begin{split}
t_1 = t(j_1, {s_1}_{\text{out}}) &\in \{4,5,\ldots , 14\} \equiv \mathcal{T}_1\\
t_2 = t(j_2, {s_1}_{\text{out}}) &\in \{1,2, \ldots,  11\} \equiv \mathcal{T}_2\\
t_1^* = t(j_1, {s_2}_{\text{out}}) &\in \{9,10,\ldots , 19\} \equiv \mathcal{T}^*_1\\
t_2^* = t(j_2, {s_2}_{\text{out}}) &\in \{10,11, \ldots,  19\} \equiv \mathcal{T}^*_2\\
t_3 = t(j_3, {s_2}_{\text{out}}) &\in \{8,9, \ldots,  18\} \equiv \mathcal{T}_3. \end{split}
\end{equation}

Now, we will investigate the linear programming approach and the time-indexed representation which leads to the QUBO formulation. For QUBO and HOBO representations, we use the following time indexed variables $x_{j_1,t_1,s_1}: t_1 \in \mathcal{T}_1$, $x_{j_1,t_1^*,s_2}: t_1^* \in \mathcal{T}^*_1$, $x_{j_2,t_2,s_1}: t_2 \in \mathcal{T}_2$, $x_{j_2,t^*_2,j_2}: t^*_2 \in \mathcal{T}^*_2$, and $x_{j_3,t_3,s_2}: t_3 \in \mathcal{T}_3$.

From Eq.~\eqref{eq:c1},
\begin{equation}\label{eq::tim}
\begin{split}
    t(j_1, {s_2}_{\text{in}}) &= t_1 + \tau^{(\text{pass})}(j_1, s_1, s_2) = t_1 + 4, \\
    t(j_2, {s_2}_{\text{in}}) &= t_2 +
    \tau^{(\text{pass})}(j_2, s_1, s_2) = t_2 + 8, \\
    t(j_3, {s_1}_{\text{in}}) &=
    t_3 + 
    \tau^{(\text{pass})}(j_3, s_2, s_1) = 
    t_3 + 8,
\end{split}
\end{equation}
and we will replace the occurrences of the variables on the left hand side using Eq.~\eqref{eq::tim} in the ILP formulation. Note that we use $t_1, t_2, t_3$ only to compute the penalty for the delays.

In the QUBO formulation, we have the following penalty term from Eq.~\eqref{eq::linear} ensuring that each train leaves each station only once:
 \begin{equation}\label{eq::toy_psum}
 \begin{split}
 &P_{\text{sum}}(\mathbf{x}) =
\\ &p_{\text{sum}} \sum_{\substack{  \mathcal{T} \in \{\mathcal{T}_1, \mathcal{T}^*_1, \mathcal{T}_2,
\mathcal{T}^*_2, \mathcal{T}_3 \} \\ (s,j) \in \{(s_1,j_1), (s_1,j_2), (s_2,j_3), (s_2,j_1), (s_2,j_2)\} }}   \left(\sum_{(t, t') \in \mathcal{T}^{\times 2},  t \neq t'} x_{j,t,s} x_{j,t',s} - \sum_{t \in \mathcal{T}} x_{j,t,s} \right).
\end{split}
\end{equation}

As the \textbf{default setting},
we consider a double-track line, where each track has its own direction (unidirectional traffic). There is a conflict between $j_1$ and $j_2$ on the line from $s_1$ to $s_2$. If $j_1$ goes first at $t=4$, then $j_2$ can start earliest at $t = 6$ (with an additional delay of $5$) to proceed at full speed. If $j_2$ goes first at $t=1$, then $j_1$ can start earliest at $t = 7$ (with an additional delay of $3$) to proceed at a full speed. In both cases, $j_3$ can proceed undisturbed. 

In the case of linear programming, the conflict can be resolved by setting the order variable $y(j_1, j_2, {s_1}_{\text out}) \in \{0,1\}$ to 1 if $j_1$ goes first and 0 if $j_2$ goes first. Recall that $y(j_2, j_1, {s_1}_{\text{out}}) = 1 - y(j_1, j_2, {s_1}_{\text{out}})$.
Referring to Eq.~\eqref{eq:c2},
\begin{equation}\label{eq::toy_linear_pair}
\begin{cases}
    t_2 + M \cdot (1- y(j_1, j_2, {s_1}_{\text{out}})) - t_1 &\geq 2 + 0, \\
        t_1 + M \cdot y(j_1, j_2, {s_1}_{\text{out}}) - t_2 &\geq 2 + 4,
        \end{cases}
\end{equation}
where $M$ is a large number. Equivalently, $t_2 - 2 < t_1 < t_2+6$ is not allowed in the time-indexed variable approach. Hence, we have the following QUBO penalty term:
\begin{equation}\label{eq::toy_QUBO_pair}
    P_{\text{pair}}^{\text{headway}}(\mathbf{x}) =  p_{\text{pair}} \sum_{t_2 - 2 < t_1 < t_2 + 6 \ t_1 \in \mathcal{T}_1,~t_2 \in \mathcal{T}_2}  x_{j_1, t_1, s_1}  x_{j_2,  t_2, s_1} + x_{j_2, t_2, s_1} x_{j_1, t_1, s_1} .
\end{equation}

We can express the minimal stay condition in Eq.~\eqref{eq:c4} as
\begin{equation}
    t_1^* -(t_1 + 4 ) \geq 1 \text{ and }
    t_2^* - (t_2 + 8) \geq  1,
\end{equation}
and the corresponding QUBO term would be

\begin{equation}
    P_{\text{pair}}^{\text{stay}}(\mathbf{x}) =  p_{\text{pair}} \left(\sum_{\substack{t_1' < t_1 + 5 \\ t_1' \in \mathcal{T}_1', t_1 \in \mathcal{T}_1}}  x_{j_1, t_1, s_{1}}  x_{j_1, t_1', s_2} +  \sum_{\substack{t_2' < t_2 + 9 \\ t_2' \in \mathcal{T}_2', t_2 \in \mathcal{T}_2}}  x_{j_2, t_2, s_{1}}  x_{j_2, t_2', s_2}
 \right).
\end{equation}
The track occupancy condition as defined in Eq.~\ref{eq:c7} for the track at platform $1$ on station $s_2$, see Fig.~\ref{fig::toy1} (both $j_1$ and $j_2$ are scheduled on this track) is expressed as
\begin{equation}
\begin{split}
 t_2 + 8 +  M \cdot (1-y(j_1, j_2, {s_2}_{\text{out}})) - t_1^* \geq 1, \\
 t_1 + 4 + M \cdot  y(j_1, j_2, {s_2}_{\text{out}} ) - t_2^*\geq  1.
\end{split}
\end{equation}
and we have
\begin{equation}\label{eq::order} y(j_1, j_2, {s_1}_{\text{out}}) = y(j_1, j_2, {s_2}_{\text{out}})
\end{equation}
as the M-P is not possible on this route (note that this last condition will be lifted while rerouting). In either case, the QUBO (HOBO) representation would be:

\begin{multline}
P_{\text{qubic}}^{\text{occ.}}(\mathbf{x}) = 2p_{\text{pair}} \left(     \sum_{\substack{t_1^* \in \mathcal{T}^*_1, \ t_2^* \in \mathcal{T}^*_2 \ t_1 \in \mathcal{T}_1  \\ t_1 +  4  - 1 < t_2^* \leq t_1^* }}  x_{j_1,t_1,s_1}  x_{j_1,t_1^*,s_2}  x_{j_2, t_2^*,s_2}\right.\\ 
\left. +\sum_{\substack{t_1^* \in \mathcal{T}^*_1, \ t_2 \in \mathcal{T}_2 \ t_2^* \in \mathcal{T}_2^*  \\ t_2 +  8  - 1 < t_1^* \leq t_2^* }}  x_{j_2,t_2,s_1}  x_{j_1,t_1^*,s_2}  x_{j_2,t_2^*,s_2}
\right),
\end{multline}
and for the decomposition we use: $ \tilde{x}_{t_1^*, t_2^*} = x_{j_1,t_1^*,s_2} \cdot x_{j_2,t_2^*,s_2}$ (where we use abbreviation $ \tilde{x}_{t_1^*, t_2^*}$ for $ \tilde{x}_{j_1, j_2, t_1^*, t_2^*, s_2}$). The first part of the qubic penalty function is given by:
\begin{multline}
P^1_{\text{qubic}}(\mathbf{x},  \mathbf{\tilde{x}} ) = p_{\text{pair}}\sum_{\substack{t_1^* \in \mathcal{T}^*_1, \ t_2^* \in \mathcal{T}^*_2 \ t_1 \in \mathcal{T}_1  \\ t_1 +  4  - 1 < t_2^* \leq t_1^* }} (x_{j_1,t_1,s_1} \tilde{x}_{t_1^*, t^*_2}  +  \tilde{x}_{t_1^*, t^*_2} x_{j_1,t_1,s_1} ) \\ +p_{\text{pair}} \sum_{\substack{t_1^* \in \mathcal{T}^*_1, \ t_2 \in \mathcal{T}_2 \ t_2^* \in \mathcal{T}_2^*  \\ t_2 +  8  - 1 < t_1^* \leq t_2^* }} (x_{j_2,t_2,s_1} \tilde{x}_{t_1^*, t^*_2}  + \tilde{x}_{t_1^*, t^*_2} x_{j_2,t_2,s_1} ).
\end{multline}
and
\begin{equation}
     P^2_{\text{qubic}}(\mathbf{x},  \mathbf{\tilde{x}} ) = p_{\text{qubic}} \sum_{t_1^* \in \mathcal{T}_1^*, t_2^* \in \mathcal{T}_2^* } h( \tilde{x}_{t_1^*, t_2^*} , \  x_{j_1,t_1^*,s_2}  , \ x_{j_2,t_2^*,s_2}),
\end{equation}
where $h$ is the polynomial from Eq.~\eqref{eq::rosenberg_polynomial}.

We use Eq.~\eqref{eq::objective} for the objective. The ILP takes the following form:

\begin{align}
    \text{min.}~~ &w_{j_1} \frac{t(j_1, {s_1}_{\text{out}}) - \nu(j_1, {s_1}_{\text{out}})}{d_{\max}} + w_{j_2} \frac{t(j_2, {s_1}_{\text{out}}) - \nu(j_2, {s_1}_{\text{out}})}{d_{\max}} \nonumber  \\ &+  w_{j_3} \frac{t(j_3, {s_2}_{\text{out}}) - \nu(j_3, {s_2}_{\text{out}})}{d_{\max}}
\label{eq::objective_linear_toy}
\end{align}
\hspace{0.4in}subject to: 
\begin{equation*}
    \begin{split}
          t_2 + M \cdot (1- y(j_1, j_2, {s_1}_{\text{out}})) - t_1 \geq  2,&  \\
        t_1 + M \cdot y(j_1, j_2, {s_1}_{\text{out}}) - t_2 \geq 6,& \\
         t_1^* - t_1 \geq  5,& \\
    t_2^* -  t_2 \geq 9,&
    \\
     t_2 + M \cdot (1-y(j_1, j_2, {s_1}_{\text{out}})) - t_1^* \geq  -7,& \\
 t_1 + M \cdot  y(j_1, j_2, {s_1}_{\text{out}} ) - t_2^* \geq  -3,& \\ 
 \\
 y(j_1, j_2, {s_1}_{\text{out}}) \in \{0, 1\},&
    \end{split}
\end{equation*}
and the range of the integer variables $t_1,t_2,t_1^*,t_2^*$ are determined by Eq.~\eqref{eq::t_limits}. We use Eq.~\eqref{eq::order} for the simplification of the precedence variables.

In QUBO formulation, we have the following objective function:
\begin{align}
    f(\mathbf{x}) =  &w_{j_1} \sum_{t \in \mathcal{T}_1}  x_{j_1,t,s_1} \frac{t - \nu(j_1,s_1)}{d_{\max}} +  w_{j_2} \sum_{t \in \mathcal{T}_2}  x_{j_2,t,s_1} \frac{t - \nu(j_2,s_1)}{d_{\max}} \nonumber \\ &+  w_{j_3} \sum_{t \in \mathcal{T}_3}  x_{j_3,t,s_2} \frac{t - \nu(j_3,s_2)}{d_{\max}}.
\label{eq::toy_objective}
\end{align}

The overall QUBO representation is expressed as

\begin{equation}\label{eg::f_default}
    \min_{\mathbf{x},  \mathbf{\tilde{x}}} q(\textbf{x},   \mathbf{\tilde{x}}) = f(\textbf{x}) + P_{\text{sum}}(\mathbf{x}) + P_{\text{pair}}^{\text{headway}}(\mathbf{x}) +  P_{\text{pair}}^{\text{stay}}(\mathbf{x})  + P^1_{\text{qubic}}(\mathbf{x},  \mathbf{\tilde{x}} ) + P^2_{\text{qubic}}(\mathbf{x},  \mathbf{\tilde{x}} ).
\end{equation}

If $j_1$ goes first, $y (j_1, j_2, {s_1}_{\text{out}}) = 1$, then we have an additional delay of $5$ from $j_2$, adding $1 \cdot \frac{5}{10} = 0.5$ to the objective. If $j_2$ goes first, $y(j_1, j_2, {s_1}_{\text{out}}) = 0$, we have an additional delay of $3$ from $j_1$, adding $2 \cdot \frac{3}{10} = 0.6$ to the objective. Therefore, at this stage, the best solution is to let $j_1$ go first, yielding $t_1 = 4$, $t_2 = 6$ and $t_3 = 8$. 

Suppose now that we find the value of the objective not satisfactory. In this case, we need to perform \textbf{rerouting}. In our case, the rerouting will concern changing the double-track line to the bidirectional traffic mode (many railway operators are being involved in such rerouting, e.g. Koleje \'Slaskie, eng. Silesian Railways). In details, there is a conflict between the trains $j_1$ and $j_2$ on the line between $s_1$ and $s_2$. Hence rerouting will be used to solve this conflict: We use the line between $s_1$ and $s_2$ as two parallel single-track lines (Track $1$ for $j_1$ and Track $2$ for $j_2$). In this case, we have no conflict between $j_1$ and $j_2$ and we lift the conditions in Eq.~\eqref{eq::toy_linear_pair} and Eq.~\eqref{eq::order} (as M-P is now possible on the line), or remove the corresponding terms from the QUBO in Eq.~\eqref{eq::toy_QUBO_pair}. However, a new conflict arises between $j_2$ and $j_3$ on the single track resource (Line $2$), so new conditions or terms will appear. Following Eq.~\eqref{eq:c3} the \textbf{single track line condition} yields: 
\begin{equation}
\begin{cases}
       t_3 + M \cdot (1-z(j_2,j_3,s_1, s_2)) - (t_2+8)\geq  1,\\
          t_2 + M \cdot z(j_2,j_3,s_1, s_2) - (t_3 + 8)\geq 1,
       \end{cases}
\end{equation}
as $\tau^{(\text{pass})}(j_2, s_1, s_2) = 8$, and $\tau^{(\text{pass})}(j_3, s_2, s_1) = 8$. Equivalently we can not have $t_3 - 8 < t_2 < t_3 + 8$) and we have the following QUBO penalty term:
\begin{equation}
    P^{1\text{track}}_{\text{pair}}(\mathbf{x}) =  p_{\text{pair}} \sum_{t_3 - 8 - 1 < t_2 < t_3 + 8 + 1 \ t_2 \in \mathcal{T}_2, t_3 \in \mathcal{T}_3}  x_{j_2, t_2, s_{1}}  x_{j_3, t_3, s_2} + x_{j_3, t_3, s_2} x_{j_2, t_2, s_{1}},
\end{equation}

The objective would be as in Eq.~\eqref{eq::objective_linear_toy}, but subject to altered constraints:
\begin{equation}
    \begin{split}
         t_3 + M \cdot (1-z(j_2,j_3,s_1, s_2)) -t_2 \geq 9,&\\
          t_2 + M \cdot z(j_2,j_3,s_1, s_2) -t_3 \geq 9,&\\
         t_1^* - t_1\geq 5,&
          \\
    t_2^* - t_2 \geq  9,&
    \\
     t_2 +  M \cdot (1 - y(j_1, j_2, {s_2}_{\text{out}})) - t_1^*\geq  -7,& \\
 t_1  + M( y(j_1, j_2, {s_2}_{\text{out}} )) - t_2^* \geq -3,& \\ 
 \\
 y(j_1, j_2, {s_2}_{\text{out}})  \in \{0, 1\},~
  z(j_2, j_3, {s_1}, s_2)  \in \{0, 1\}, &
          \\
    \end{split}
\end{equation}
and the ranges of the integer variables $t_1,t_2,t_1^*,t_2^*$ are determined by Eq.~\eqref{eq::t_limits}.

The effective QUBO representation here is given as
\begin{equation}\label{eg::f_rerout}
     \min_{\mathbf{x}, \mathbf{\tilde{x}}} q^r(\mathbf{x}, \mathbf{\tilde{x}})
 =  f(\mathbf{x}) + P_{\text{sum}}(\mathbf{x}) +  P_{\text{pair}}^{\text{stay}}(\mathbf{x}) + P_{\text{pair}}^{1\text{track}}(\mathbf{x}) + P^1_{\text{qubic}}(\mathbf{x},  \mathbf{\tilde{x}} ) + P^2_{\text{qubic}}(\mathbf{x},  \mathbf{\tilde{x}} ).
\end{equation}

If $j_3$ goes first ($z(j_2,j_3,s_1, s_2) = 0$), the additional delay of $j_2$ would exceed the maximal $d_{\text{max}} = 10$. The optimal solution is $z(j_2,j_3,s_1, s_2) = 1$ and $y(j_1, j_2, {s_2}_{\text{out}}) = 1$, hence $t_1 = 4$, $t_2 = 2$,  $t_3 = 11$, and $t_1^* = 9$.
The additional delay of $j_1$ is $0$, $j_2$ is $1$, and $j_3$ is $3$ with the objective $0.4$, which is better than the objective of the default settings. As there is no possibility to reroute trains further to lift the conflict between $j_2$ and $j_3$, we can consider this objective as the optimal one.

\subsection{Numerical calculations}\label{sec::calculations_small}

In this section, we present a proof-of-concept by solving the small numerical example described above using D-Wave solvers.  

We first solved the problem using the ILP formulation to test the validity of the model. We used Python $3.9$ programming language and PulP library \cite{Mitchell11pulp:a} to implement the ILP formulation and CBC (Coin-or branch and cut)\cite{forrest2005cbc} solver to solve the problem, which is the default solver in PulP, to test the validity of the model. We reached $t_1 = 4$, $t_2 = 6$ and $t_3 = 8$ for the default settings (with objective $0.5$), see Tab.~\ref{tab::results_linear_default}, and $t_1 = 4$, $t_2 = 2$ and $t_3 = 11$ for the rerouting (with objective $0.4$) as expected, see Tab.~\ref{tab::results_linear_rerout}. Note that we are not interested in the run-time comparison between the linear solver and D-Wave, but we would like to demonstrate the potential of quantum annealing for solving train rescheduling problems.

\begin{table}[ht]
\centering
\subfloat[Default Settings.]{
\begin{tabular}{lcccc}
\hline
\textbf{}      & \textbf{$S_1$} & \multicolumn{2}{c}{\textbf{$S_2$}} & \textbf{$S_1$} \\ \hline
\textbf{}      & \textbf{Dep.}  & \textbf{Arr.}    & \textbf{Dep.}   & \textbf{Arr.}  \\ \hline
\textbf{$j_1$} & 4              & 8                & 9               &                \\ \hline
\textbf{$j_2$} & 6              & 14               &                 &                \\ \hline
\textbf{$j_3$} &                &                  & 8               & 16             \\ \hline
\end{tabular}\label{tab::results_linear_default}
}
\quad
\subfloat[Rerouting.]{
\begin{tabular}{lcccc}
\hline
\textbf{}      & \textbf{$S_1$} & \multicolumn{2}{c}{\textbf{$S_2$}} & \textbf{$S_1$} \\ \hline
\textbf{}      & \textbf{Dep.}  & \textbf{Arr.}    & \textbf{Dep.}   & \textbf{Arr.}  \\ \hline
\textbf{$j_1$} & 4              & 8                & 9               &                \\ \hline
\textbf{$j_2$} & 2              & 10               &                 &                \\ \hline
\textbf{$j_3$} &                &                  & 11               & 19             \\ \hline
\end{tabular}\label{tab::results_linear_rerout}
}
\caption{Solutions obtained from the linear solver.}
\end{table}

 We implemented the QUBO formulation presented in Section 3.3 using D-Wave Ocean SDK. For the numerical calculations on the D-Wave machine we need to pick particular penalty values. The theory of penalty methods is discussed, for example, in \cite{luenberger2015linear}. In general, the solution of the unconstrained objective tends to be a feasible optimal solution to the original problem as the penalties of constraints tend to infinity. However, in practice, these penalties have to be chosen so that the constraint terms do not dominate over the objective. If the penalties are too high, the objective can be lost in the noise of the physical quantum annealer. Based on these heuristics, we used the following strategy in the determination of penalties: 
\begin{enumerate}
    \item Penalties for the hard constraints ($p_{\text{sum}}$, $2 p_{\text{pair}}$, and $p_{\text{qubic}}$) should be higher than the maximal possible objective for the single train, i.e. $\max_{j} w_j = 2$
    \item $p_{\text{qubic}}$ should be smaller than other hard penalties, as it is multiplied by $3$ or $2$ in some terms of HOBO - see Eq.~\eqref{eq::rosenberg_polynomial}.
    \item We pick penalties of hard constraints as low as possible, to prevent the objective from being overriden by the noise of the quantum annealer.
\end{enumerate}
The terms in Eq.~\eqref{eq::toy_objective} (the maximal penalty here is $w_{j_1} = 2.$) are ``soft constraints'', and the terms in Eq.~\eqref{eg::f_default} and Eq.~\eqref{eg::f_rerout} are the ``hard constraints'' that can not be broken for the solution to be feasible. 
Hence, we use the following penalty parameters $p_{\text{sum}} = 2.5$, $p_{\text{pair}} = 1.25$ (as each element is taken twice) and $p_{\text{qubic}} = 2.1$. 

Both for the default settings and rerouting, we had $176$ logical variables, out of which $55$ were the $\mathbf{x}$ variables and $121$ were the auxiliary $ \mathbf{\tilde{x}}$ variables. Here we have a relatively large overhead due to the cubic term. Hence the single track occupation condition has to be used with caution when handling large railway problems. To test the validity of the model, we first solved the two problems using the simulated annealer (SA) from the D-Wave Ocean SDK, which is a classical heuristic algorithm for solving combinatorial optimization problems stated as QUBOs. When running SA or QA, the output is a list of samples (0-1 assignments to the binary variables) and the corresponding energies (value of $q(x)$). The lowest energy solution is called the ground state. Using SA, We got the same solutions as the linear solver with the following energies $q(\textbf{x},   \mathbf{\tilde{x}}) = -12.0$ and $q^r(\textbf{x},   \mathbf{\tilde{x}}) = -12.1$. The energies correspond to the ground state as $-12.5$ is the offset (the constant term in the QUBO formulation ), and $0.5$ and $0.4$ are the optimal (lowest possible) penalties for delays.  

Next, we solved the problem on D-Wave Advantage quantum processing unit (QPU) \cite{johnson2011quantum}. In D-Wave Advantage QPU, not all the qubits are interconnected via couplers, and the underlying graph has the specific structure known as the Pegasus topology \cite{dattani2019pegasus}. Hence, before running a problem on the D-Wave, a procedure called \emph{minor embedding} is required to map the logical variables to the physical qubits on the machine. Due to limited connectivity, a single logical qubit is often represented with a \emph{chain} of physical qubits that are coupled strong enough so that they end up in the same value representing the same variable. The coupling between the qubits in the chain is known as the \emph{chain strength}, and a low chain strength may result in chain breaks while a high chain strength may override the problem parameters. In our experiments on D-Wave Advantage, we used the default minor embedding algorithm provided by Ocean SDK and used various chain strengths. The number of logical variables is 176 and the number of physical qubits used in the machine after embedding is $\sim 900$. For both problems the degree of completeness of the problem graph was approximately $0.1$.

\begin{figure}
    \centering
    \includegraphics[width=\linewidth]{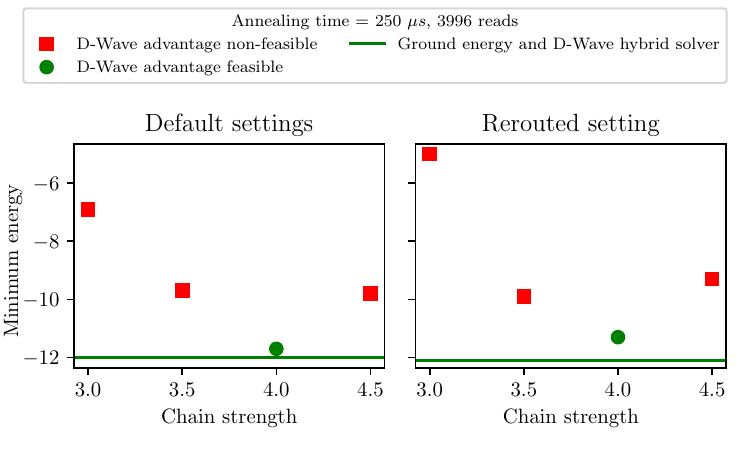}
    \caption{Lowest energy solutions obtained from D-Wave Advantage and D-Wave hybrid solvers. In the case of D-Wave Advantage, only one feasible solution was found for each panel. For chain strength $4.0$ at each panel, the percentage of feasible solutions over total number of solutions is roughly $2.5 \times 10^{-4}$.}
    \label{fig::D_Wave}
\end{figure}

Another parameter that needs to be set is the annealing time. Annealing time depends on the problem and problem size and is also limited by the current technology of D-Wave Advantage QPU. In our experiments, the annealing time is set as $250 \mu s$. Results of the D-Wave experiments are presented in Fig.~\ref{fig::D_Wave}. A solution is feasible, if it can be technically realized on the railroad infrastructure, i.e., all hard constraints are fulfilled. A solution is optimal if the order of the trains on conflicting resources (i.e., tracks that are used by more than one train) is the same as the order in the ground state solution. We reached optimal solutions (in the sense of the train order) using the D-Wave machine, both for the default settings and rerouting.

For the default settings, D-Wave results for chain strength 4 are: $t_1 = 4$, $t_2 = 8$ (adding an additional $0.7$ to the objective), $t_3 = 9$ (adding an additional $0.1$ to the objective) and $t_1^* = 10$, see Tab.~\ref{tab::results_dw_default}. The solution is feasible, since $j_2$ leaves $s_1$ at $t_2 = 8$, late enough to have no conflict with $j_1$ which will leaves $s_1$ at $t_1 = 4$. Furthermore, $j_2$ will arrive to $s_2$ at $t_2 + \tau^{(pass)}(j_2, s_1, s_2) = 8 + 8 = 16$, i.e. after $j_1$ leaves $s_2$ at $t_1' = 11$. The order of trains is the same as in the optimal solution, and the energy of the state is $-11.7$. This energy does not correspond to the ground state as there are some additional delays of the trains, however, do not affect the feasibility and the order of trains.

For rerouting, the results of D-Wave are: $t_1 = 6$, (adding an additional $0.4$ to the objective) $t_2 = 4$ (adding an additional $0.3$ to the objective), $t_3 = 13$ (adding an additional $0.5$ to the objective) and $t_1^* = 11$, see Tab.~\ref{tab::results_dw_rerout}. The solution is feasible, since $j_2$ will arrive to $s_2$ at $t_2 + \tau^{(pass)}(j_2, s_1, s_2) = 4 + 8 = 12$, i.e. after $j_1$ leaves $s_2$ at $t_1^* = 11$, and before $j_3$ leaves $s_2$ at $t_3 = 13$. The order of trains is the same as in the optimal solution, the state energy is $-11.3$. Like in the default settings case, the found solution is feasible but not the ground state.

\begin{table}[ht]
\centering
\subfloat[Default Settings]{
\begin{tabular}{lcccc}
\hline
\textbf{}      & \textbf{$S_1$} & \multicolumn{2}{c}{\textbf{$S_2$}} & \textbf{$S_1$} \\ \hline
\textbf{}      & \textbf{Dep.}  & \textbf{Arr.}    & \textbf{Dep.}   & \textbf{Arr.}  \\ \hline
\textbf{$j_1$} & 4              & 8                & 10               &                \\ \hline
\textbf{$j_2$} & 8              & 16               &                 &                \\ \hline
\textbf{$j_3$} &                &                  & 9               & 17             \\ \hline
\end{tabular}\label{tab::results_dw_default}
}
\quad
\subfloat[Rerouting]{
\begin{tabular}{lcccc}
\hline
\textbf{}      & \textbf{$S_1$} & \multicolumn{2}{c}{\textbf{$S_2$}} & \textbf{$S_1$} \\ \hline
\textbf{}      & \textbf{Dep.}  & \textbf{Arr.}    & \textbf{Dep.}   & \textbf{Arr.}  \\ \hline
\textbf{$j_1$} & 6              & 10                & 11               &                \\ \hline
\textbf{$j_2$} & 4              & 12               &                 &                \\ \hline
\textbf{$j_3$} &                &                  & 13               & 21             \\ \hline
\end{tabular}\label{tab::results_dw_rerout}
}
\caption{Solutions obtained from the D-Wave.}
\end{table}

Another alternative is to use the hybrid solver for binary quadratic models provided by D-Wave. The hybrid solver runs in parallel modules consisting of a heuristic classical component to explore the search space and a quantum component that makes queries to D-Wave QPU to guide the optimization process and improve the existing solutions. The best solution found among the parallel runs is returned to the user \cite{hybrid}. Using the hybrid solver, we obtained the ground state, both in the case of default settings and the rerouted setting.

With our example, we have demonstrated that although it is possible to have the optimal solution for the D-Wave, it is not straightforward and requires at least an extensive parameter sweep. On the other hand, the D-Wave hybrid solver found the ground state on the first try. More importantly, the hybrid solver can be used for tackling larger problems as those solvers can work on problem instances with up to 20000 variables that are fully connected or up to 1 million variables for sparse $Q$ matrices \cite{mcgeoch2020d}.

\subsection{Assesment of solvers on larger instances}

To demonstrate the feasibility of the hybrid solver, we have assessed both the D-Wave Advantage and the D-Wave hybrid solver on a bit larger examples. In both examples, we use the same parameters settings as in Sec.~\ref{sec::calculations_small} for the calculations.

The first is a bit enlarged \textbf{default setting} one with infrastructure as in Fig.~\ref{fig::toy1}. Here, in addition, train $j_3$ is followed by another stopping train $j_4$, and the conflict occurs on the \textbf{minimal headway} between $j_3$ and $j_4$. We call the problem the \textbf{4 trains 2 stations} example. The problem is a bit larger with $187$ logical variables. Although the number of connections is larger, the degree of completeness of the graph is a bit smaller and equals roughly $0.09$. The ground state energy, consistent with the solution of the ILP, equals to $q(\textbf{x}, \mathbf{\tilde{x}}) = -14.4$.

The second example concerns a larger number of trains and a larger number of stations on a more branched network. We call the problem the \textbf{5 trains 5 stations} example. The problem is encoded on $341$ logical variables, but with a much smaller degree of completeness of the graph which equals roughly $0.04$.
The ground state energy, consistent with the solution of the ILP, equals to $q(\textbf{x}, \mathbf{\tilde{x}}) = -21.49$.

\begin{figure}[t!]
    \centering
    \includegraphics[width=\linewidth]{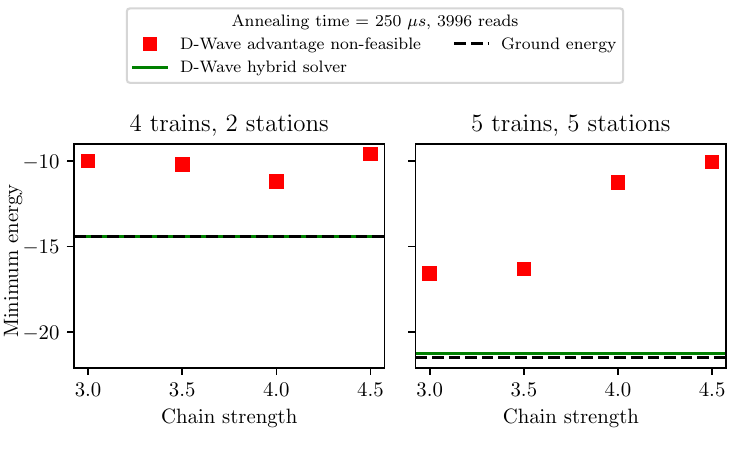}
    \caption{Lowest energy solutions obtained from D-Wave Advantage and D-Wave hybrid solvers for larger problems. We present also comparison with the ground state achieved from the ILP (classical) approach.}
    \label{fig::D_Wave_5t}
\end{figure}

Results of calculations for both additional examples are presented in Fig.~\ref{fig::D_Wave_5t}.
As we can see for slightly larger problems than in Sec.~\ref{sec::calculations_small}, the D-Wave Advantage does not give any feasible solution. The D-Wave hybrid solver, on the other hand, still has promising outcomes.
Actual characteristics of the problem are presented in Tab.~\ref{tab::nvars}.
Here, we have observed that the larger the railway problem is, the smaller the degree of completeness. This observation coincides with Tab. IV ~\cite{domino2020quantum} and discussion in Sec.~\ref{sec:prob} as the number of variables and number of non-zero QUBO terms are roughly linear in the number of trains and stations.
Referring to Fig.~\ref{fig::D_Wave}, Fig.~\ref{fig::D_Wave_5t}, and Tab. IV in~\cite{domino2020quantum} we can generally conclude that smaller railway problems, with the graph's degree of completeness of $0.1$ or larger, are solvable on the D-Wave machine without the need for the D-Wave hybrid solver. For larger problems, the hybrid solver is necessary.

\begin{table}[ht]
\centering
\begin{tabular}{lcccc}
\hline
\textbf{}     & \multicolumn{4}{c}{Problem}  \\ \hline
\textbf{}      & \textbf{default}  & \textbf{rerouted}    & \textbf{4 trains 2 stations}   & \textbf{5 trains 5 stations}  \\ \hline
\# logical vars. & 176              & 176                & 187              &  341              \\ \hline
\# edges &    1492            &   1504               & 1577               & 2159             \\ \hline
density vs. full  &    0.097           &   0.098               & 0.091               & 0.037            \\ \hline

\# physical vars. &   869            & 739                 &      979           &     1419           \\ \hline
\end{tabular}
\caption{Problem characteristics. The first $3$ rows refer to the characteristics of the QUBO and the last one to the characteristics of the solver. Number of physical variables is computed via D-Wave's default embedding algorithm minorminer which is a heuristic algorithm, hence the data are approximate.}
\label{tab::nvars}
\end{table}

From a practical point of view, the above-presented problems are still of small size due to the small size of the current D-Wave machine. To estimate the amount of logical resources needed to solve real-life problems, let us consider an hour cycle on the dense traffic (one train per $2$ min. in each direction) on the double-track metro line with $20$ stations. (In an hour cycle, we have $60$ trains.) We then consider $d_{\text{max}} = 5$ minutes, and $1$ minute resolution. According to Eq.~(49), we would have roughly $7\_200$ variables. If each train is assumed to be in potential conflict with $H(d_{\text{max}}) = 5$ other trains (that many trains pass in $2 d_{\text{max}} = 10$ min. interval), then according to Eq.~(55) we will have roughly $216\_000$ auxiliary variables. (Obviously, in both cases, the particular number of variables depends on the details of the topology of the problem.) Such a problem would be solvable on the not very large device but with possible $3$rd order connections or a much larger one with $2$nd order connections only.

\section{Conclusions and outlook}

As current classical technologies are insufficient to compute feasible results in a reasonable time, fully developed quantum annealing offers potential polynomial speed-ups for these problems. However, switching from the conventional notation to the one demanded by the quantum annealer is a challenge. Our paper is the first to present the quadratic and higher-order unconstrained binary optimization representation for the railway rescheduling problem concerning the determination of the order of trains on conflicted resources, rescheduling, and rerouting trains on single-, double- and multi-track lines and stations.

The number of qubits is one of the bottlenecks of current quantum devices. It is thus desirable to use the smallest possible number of qubits when modeling. When quadratic and higher-order models are compared, the latter is more efficient in terms of the number of qubits required. Although currently, it is not possible to utilize HOBO with quantum annealers, the need for quantum annealers allowing such interactions is evident \cite{perdomo2019readiness}. There is also ongoing work for building architectures that allow solving optimization problems involving higher-order terms directly \cite{ender2021parity} in the gate-based model. Furthermore, algorithms like quantum approximate optimization algorithm (QAOA) \cite{farhi2014quantum} allow solving higher-order problems natively \cite{glos2020space,tabi2020quantum}.


Four demonstrative problems were implemented on the current D-Wave machine. Two smaller problems were successfully solved both using the D-Wave Advantage QPU and using the D-Wave hybrid solver. Two larger problems were successfully solved only on D-Wave hybrid solver which we find promising for solving larger instances. Importantly, we have presented the HOBO/QUBO formulation that can be used with quantum-inspired architectures designed for solving combinatorial optimization problems stated in QUBO form such as Fujitsu digital annealers \cite{tsukamoto2017accelerator}.

Determination of penalty values poses a challenge for solving QUBO problems in general. Although we have determined the penalty values using heuristic methods, note that there are some recent algorithms dedicated to penalty determination like the cross entropy optimization discussed in \cite{roch2021cross} and the one discussed in \cite{ayodele2022multi} (see. Eg. Section 3.2) is tested successfully on the particular Fujitsu digital annealer.

Curiosity arises on how quantum annealers or other Ising-based heuristics behave in solving real-life problems compared to conventional methods. Further research should be undertaken to explore the applicability of the presented approach for real-life train rerouting and rescheduling problems. In particular, when considering the railway traffic on the regional scale where delays can be large and the number of trains is not very large, the QUBO formulation that will be obtained from the ILP representation presented in this paper may be worth investigating.

Besides wide railway potential applications (ordinary railways, metro, trams), discussed rules of problem conversion into HOBO / QUBO can be applied generically in many branches of operational research. Let us list a few:
\begin{enumerate}
    \item Electric bus scheduling, where the charging place occupation condition can be modeled in analogy to our track occupation condition.
    \item Automated guided vehicle (AGV) scheduling in the factory, where there are many railway analogies. AGVs have a pre-designed schedule that is conflicted and needs to be rescheduled. AGVs follow the paths that are uni or bi-directional; hence, there is a headway and single track line condition. There are places that can be occupied by one AGV at a time (track occupation condition), paths of AGVs cross (switch condition), and there is the sequence of tasks for the given trolley (rolling stock circulation condition). Rerouting of AGVs can be treated as an extra task beyond the optimization as in Fig. 4., and finally, AGVs may have various priorities.
\end{enumerate}
In general, our HOBO approach (generated by \textbf{track occupation condition}) may be applicable for models consisting of ''stations'' that can be occupied by only one ''vehicle'' at a time, with waiting possibilities on stations and no-waiting elsewhere.

\section*{Data availability}
The code and the data used for generating the numerical results can be found in \url{https://github.com/iitis/railways\_HOBO}.

\section*{Acknowledgement} 
The research was supported by the Foundation for Polish Science (FNP) under grant number TEAM NET POIR.04.04.00-00-17C1/18-00 (KD); and the National
Science Centre (NCN), Poland, under project number 2019/33/B/ST6/02011 (A.K. and \"O.S.) and by the Silesian University
of Technology Rector’s Grant no.  BKM-700/RT2/2022\ \  12/020/BKM2022/0233 (KK). We acknowledge the cooperation with Koleje \'Slaskie sp. z o.o. (eng. Silesian Railways) and appreciate the valuable and substantive discussions. We acknowledge the
consultation with company Aiut ltd. (Wyczółkowskiego street Gliwice) on the rescheduling/rerouting of industrial trolleys (AGVs).

\bibliography{quantum_silesia}

\begin{appendix}

\section{Appendix - railway terminology}
\label{appendix:: definitions-rail-term} 

For the clarity of presentation for the non-railway community, we include  the clear definitions of railway terminology in the form of Table~\ref{tab:railways-terminology-definition}.
\begin{table}[tbh!]
\centering
\begin{tabular}{lp{0.58\textwidth}}
\textbf{Railway terms} & \textbf{Definitions} \\ \hline
Schedule                            &    Pre-set sequences of blocks with departure times assigned to selected station blocks                             \\ \hline
Station blocks                         & Track section between two signalling utilities at stations.\\ \hline
Line blocks                         & Track section between two signalling utilities on lines.\\ \hline
Rerouting          &    Change of the track used by a train within a line or a station                              \\ \hline
Rescheduling           &    Train departure time modification in a way to avoid conflict                             \\ \hline
Meet and pass (M-P)                 &  Meeting of two trains at the same spatial location while following the same route in opposite directions. \\ \hline
Meet and overtake (M-O)             &  Overtaking one train by another while following the same route in the same direction. \\ \hline
Unavoidable delay                   &     Delay from outside the model that is propagated through the network, not including any delay that may be caused by other trains' traffic                            \\ \hline
Additional delay                     &     Delay beyond unavoidable caused by solving conflicts due to traffic, that is in control of our model                             \\ \hline
Headway                             &    Minimal time span between trains. \\ \hline
\end{tabular}
\caption{Summary of definitions of railway terminologies.}
\label{tab:railways-terminology-definition}
\end{table}

\end{appendix}

\end{document}